\documentclass[twocolumn,10pt]{article}
\usepackage[utf8]{inputenc}    
\usepackage{amsmath,amssymb}   
\usepackage{graphicx}          
\usepackage{authblk}           
\usepackage{lineno}            
\usepackage{rotating}          
\usepackage{subfigure}         
\usepackage{xspace}            
\usepackage{color, xcolor}     
\usepackage{titlesec}          
\usepackage{fancyhdr}          
\usepackage{float}             
\usepackage[top=3cm, bottom=3cm, left=2cm, right=2cm]{geometry} 
\usepackage[colorlinks]{hyperref} 
\usepackage[numbers,sort]{natbib} 
\graphicspath{{images/}}        
\DeclareGraphicsExtensions{.pdf,.jpeg,.png,.eps} 
\usepackage{epstopdf}           
\definecolor{faintgray}{RGB}{10,10,10} 
\titleformat{\subsection}
  {\color{faintgray}\normalfont\normalsize\bfseries} 
  {\thesubsection}{1em}{}                       
\usepackage[labelfont=bf, labelsep=colon, font=small]{caption}

\usepackage{csquotes}
\title{\textbf{Examining hadronic resonance dynamics at energies available at the CERN Large Hadron Collider: Insights from EPOS4}}
\author[1]{\textbf{Vikash Sumberia}\thanks{Corresponding author: \textit{vikash.sumberia@cern.ch}}}
\author[2]{\textbf{Dukhishyam Mallick}\thanks{Corresponding author: \textit{dukhishyam.mallick@cern.ch}}}
\author[1]{Sanjeev Singh Sambyal\thanks{\textit{sanjeev.singh.sambyal@cern.ch}}}
\author[1]{Nasir Mehdi Malik\thanks{\textit{nasir.mehdi.malik@cern.ch}}}

\affil[1]{University of Jammu, J\&K, India}
\affil[2]{Université Paris-Saclay, CNRS/IN2P3, IJCLab, Orsay, France}

\begin{document}
%

\newcommand{\pp}           {pp\xspace}
\newcommand{\ppbar}        {\mbox{$\mathrm {p\overline{p}}$}\xspace}
\newcommand{\XeXe}         {\mbox{Xe--Xe}\xspace}
\newcommand{\PbPb}         {\mbox{Pb--Pb}\xspace}
\newcommand{\pA}           {\mbox{pA}\xspace}
\newcommand{\AAa}          {\mbox{AA}\xspace}
\newcommand{\pPb}          {\mbox{p--Pb}\xspace}
\newcommand{\AuAu}         {\mbox{Au--Au}\xspace}
\newcommand{\dAu}          {\mbox{d--Au}\xspace}
\newcommand{\CuCu}         {\mbox{Cu--Cu}\xspace}

\newcommand{\s}            {\ensuremath{\sqrt{s}}\xspace}
\newcommand{\snn}          {\ensuremath{\sqrt{s_{\mathrm{NN}}}}\xspace}
\newcommand{\pt}           {\ensuremath{p_{\rm T}}\xspace}
\newcommand{\meanpt}       {$\langle p_{\mathrm{T}}\rangle$\xspace}
\newcommand{\ycms}         {\ensuremath{y_{\rm CMS}}\xspace}
\newcommand{\ylab}         {\ensuremath{y_{\rm lab}}\xspace}
\newcommand{\etarange}[1]  {\mbox{$\left | \eta \right |~<~#1$}}
\newcommand{\etaran}[2]    {#1  $< \eta <$  #2}
\newcommand{\yrange}[1]    {\mbox{$\left | y \right |<$~0.5}}
\newcommand{\yran}[2]      {#1 $< y_{\rm{cm}} <$ #2}
\newcommand{\dndy}         {\ensuremath{\mathrm{d}N_\mathrm{ch}/\mathrm{d}y}\xspace}
\newcommand{\dndeta}       {\ensuremath{\mathrm{d}N_\mathrm{ch}/\mathrm{d}\eta}\xspace}
\newcommand{\avdndeta}     {\ensuremath{\langle\dndeta\rangle}\xspace}
\newcommand{\dNdy}         {\ensuremath{\mathrm{d}N_\mathrm{ch}/\mathrm{d}y}\xspace}
\newcommand{\dNdyy}        {\ensuremath{\mathrm{d}N/\mathrm{d}y}\xspace}
\newcommand{\Npart}        {\ensuremath{N_\mathrm{part}}\xspace}
\newcommand{\Ncoll}        {\ensuremath{N_\mathrm{coll}}\xspace}
\newcommand{\dEdx}         {\ensuremath{\textrm{d}E/\textrm{d}x}\xspace}
\newcommand{\RpPb}         {\ensuremath{R_{\rm pPb}}\xspace}
\newcommand{\RAA}         {\ensuremath{R_{\rm AA}}\xspace}
\newcommand{\RpA}         {\ensuremath{R_{\rm pA}}\xspace}
\newcommand{\res}         {\ensuremath{\sigma}\xspace}

\newcommand{\nineH}        {$\sqrt{s}~=~0.9$~Te\kern-.1emV\xspace}
\newcommand{\seven}        {$\sqrt{s}~=~7$~Te\kern-.1emV\xspace}
\newcommand{\twoH}         {$\sqrt{s}~=~0.2$~Te\kern-.1emV\xspace}
\newcommand{\twosevensix}  {$\sqrt{s}~=~2.76$~Te\kern-.1emV\xspace}
\newcommand{\five}         {$\sqrt{s}~=~5.02$~Te\kern-.1emV\xspace}
\newcommand{\twosevensixnn}{$\sqrt{s_{\mathrm{NN}}}~=~2.76$~Te\kern-.1emV\xspace}
\newcommand{\fivenn}       {$\sqrt{s_{\mathrm{NN}}}~=~5.02$~Te\kern-.1emV\xspace}
\newcommand{\LT}           {L{\'e}vy-Tsallis\xspace}
\newcommand{\GeVc}         {Ge\kern-.1emV/$c$\xspace}
\newcommand{\MeVc}         {Me\kern-.1emV/$c$\xspace}
\newcommand{\TeV}          {Te\kern-.1emV\xspace}
\newcommand{\GeV}          {Ge\kern-.1emV\xspace}
\newcommand{\MeV}          {Me\kern-.1emV\xspace}
\newcommand{\GeVmass}      {Ge\kern-.2emV/$c^2$\xspace}
\newcommand{\MeVmass}      {Me\kern-.2emV/$c^2$\xspace}
\newcommand{\lumi}         {\ensuremath{\mathcal{L}}\xspace}
\newcommand{\fmc}          {fm/$c$\xspace}

\newcommand{\ITS}          {\rm{ITS}\xspace}
\newcommand{\TOF}          {\rm{TOF}\xspace}
\newcommand{\ZDC}          {\rm{ZDC}\xspace}
\newcommand{\ZDCs}         {\rm{ZDCs}\xspace}
\newcommand{\ZNA}          {\rm{ZNA}\xspace}
\newcommand{\ZNC}          {\rm{ZNC}\xspace}
\newcommand{\SPD}          {\rm{SPD}\xspace}
\newcommand{\SDD}          {\rm{SDD}\xspace}
\newcommand{\SSD}          {\rm{SSD}\xspace}
\newcommand{\TPC}          {\rm{TPC}\xspace}
\newcommand{\TRD}          {\rm{TRD}\xspace}
\newcommand{\VZERO}        {\rm{V0}\xspace}
\newcommand{\VZEROA}       {\rm{V0A}\xspace}
\newcommand{\VZEROC}       {\rm{V0C}\xspace}
\newcommand{\Vdecay} 	   {\ensuremath{V^{0}}\xspace}

\newcommand{\ee}           {\ensuremath{e^{+}e^{-}}} 
\newcommand{\pip}          {\ensuremath{\pi^{+}}\xspace}
\newcommand{\pim}          {\ensuremath{\pi^{-}}\xspace}
\newcommand{\pipm}         {\ensuremath{\pi^{\pm}}\xspace}
\newcommand{\kap}          {\ensuremath{\rm{K}^{+}}\xspace}
\newcommand{\kam}          {\ensuremath{\rm{K}^{-}}\xspace}
\newcommand{\pbar}         {\ensuremath{\rm\overline{p}}\xspace}
\newcommand{\kzero}        {\ensuremath{{\rm K}^{0}_{\rm{S}}}\xspace}
\newcommand{\lmbmass}      {\ensuremath{\Lambda(1115)}\xspace}
\newcommand{\lmb}          {\ensuremath{\rm{\Lambda}}\xspace}
\newcommand{\almb}         {\ensuremath{\overline{\rm{\Lambda}}}\xspace}
\newcommand{\Om}           {\ensuremath{\Omega^-}\xspace}
\newcommand{\Mo}           {\ensuremath{\overline{\Omega}^+}\xspace}
\newcommand{\X}            {\ensuremath{\Xi^-}\xspace}
\newcommand{\Ix}           {\ensuremath{\overline{\Xi}^+}\xspace}
\newcommand{\Xis}          {\ensuremath{\Xi^{\pm}}\xspace}
\newcommand{\Oms}          {\ensuremath{\Omega^{\pm}}\xspace}
\newcommand{\degree}       {\ensuremath{^{\rm o}}\xspace}
\newcommand{\rh}           {\ensuremath{\rm {\rho}^{\rm 0}}\xspace}
\newcommand{\rhmass}       {\ensuremath{\rm {\rho(770)}^{\rm 0}}\xspace}
\newcommand{\kstarmass}    {\ensuremath{\rm {K^{*}(892)}^{\rm{ 0}}}\xspace}
\newcommand{\kstarpmmass}  {\ensuremath{\rm {K(892)}^{\rm{*\pm}}}\xspace}
\newcommand{\kstar}        {\ensuremath{\rm {K}^{\rm{* 0}}}\xspace}
\newcommand{\kstarpm}      {\ensuremath{\rm {K}^{\rm{*\pm}}}\xspace}
\newcommand{\sigmass}      {\ensuremath{\rm {\Sigma(1385)}^{\rm{\pm}}}\xspace}
\newcommand{\sigm}         {\ensuremath{\rm {\Sigma}^{\rm{\pm}}}\xspace}
\newcommand{\sigmstar}     {\ensuremath{\rm {\Sigma}^{\rm{*\pm}}}\xspace}
\newcommand{\ximass}       {\ensuremath{\rm {\Xi(1530)}^{\rm{0}}}\xspace}
\newcommand{\xim}          {\ensuremath{\rm {\Xi}^{\rm{0}}}\xspace}
\newcommand{\ximinus}      {\ensuremath{\rm {\Xi}^{\rm{-}}}\xspace}
\newcommand{\xistar}      {\ensuremath{\rm {\Xi}^{*0}}\xspace}
\newcommand{\lambmass}     {\ensuremath{\rm {\Lambda(1520)}}\xspace}
\newcommand{\alambmass}     {\ensuremath{\rm {\Bar{\Lambda}(1520)}}\xspace}
\newcommand{\lstar}        {\ensuremath{\rm {\Lambda}^{\rm{* }}}\xspace}
\newcommand{\delstar}        {\ensuremath{\rm {\Delta}^{\rm{++ }}}\xspace}
\newcommand{\phim}         {\ensuremath{\phi}\xspace}
\newcommand{\phimass}      {\ensuremath{\phi(1020)}\xspace}
\newcommand{\pik}          {\ensuremath{\pi\rm{K}}\xspace}
\newcommand{\kk}           {\ensuremath{\rm{K}\rm{K}}\xspace}
\newcommand{\zero}         {\ensuremath{^{\rm 0}}\xspace}
\newcommand{\kskm}{$\mathrm{K^{*0}/K^{-}}$}
\newcommand{\phikm}{$\mathrm{\phi/K^{-}}$}
\newcommand{\phixi}{$\mathrm{\phi/\Xi}$}
\newcommand{\phiom}{$\mathrm{\phi/\Omega}$}
\newcommand{\xiphi}{$\mathrm{\Xi/\phi}$}
\newcommand{\omphi}{$\mathrm{\Omega/\phi}$}
\newcommand{\kstf} {K$^{*}(892)^{0}~$}
\newcommand{\phf} {$\mathrm{\phi(1020)}~$}
\newcommand{\dd}{\ensuremath{\mathrm{d}}}
\newcommand{\mT}{\ensuremath{m_{\mathrm{T}}}\xspace}
\newcommand{\krr}{\ensuremath{\kern-0.09em}}
\newcommand{\npart}{\ensuremath{\langle N_{\mathrm{part}}\rangle}\xspace}
\newcommand{\ncoll}{\ensuremath{\langle N_{\mathrm{coll}}\rangle}\xspace}

\bibliographystyle{apsrev}
\maketitle
\thispagestyle{empty} 

\begin{abstract}
\sloppy
Hadronic resonances, with lifetimes of a few \fmc are key tools for studying the hadronic phase in high-energy collisions. This work investigates resonance production in pp collisions at \s = 13.6 \TeV and in Pb--Pb collisions at \snn = 5.36 \TeV using the EPOS4 model. The EPOS4 model provides the capability to switch the Ultra-Relativistic Quantum Molecular Dynamics (UrQMD) ON or OFF, allowing the study of final-state hadronic interactions in the hadronic phase. Although this study is primarily focused on understanding the production of hadronic resonances, we also investigate the production of  strange and non-strange hadrons in relation to rescattering, regeneration, baryon-to-meson production, and strangeness enhancement, using transverse momentum (\pt) spectra and particle ratios.

Rescattering effects and strangeness enhancement play an important role in the low-\pt region, where soft, non-perturbative processes dominate particle production. At intermediate-\pt, the baryon-to-meson  yield ratios show an enhancement, and the strong mass-dependent radial flow observed for baryons in the most central \PbPb collisions  reflects the interplay between hadronization dynamics and collective effects.
The average \pt scaled with the reduced hadron mass (i.e., the mass divided by the number of valence quarks) is examined, revealing a deviation from a linear trend for short-lived resonances, attributed to hadronic phase effects. By analyzing the yield ratios of short-lived resonances to stable hadrons in pp and \PbPb collisions, the time duration ($\tau$) of the hadronic phase as a function of average charged multiplicity is estimated. The results show that $\tau$ increases with increasing multiplicity and system size, also exhibiting a non-zero value for high-multiplicity pp collisions. It is found that the production of baryons such as non-strange (p), strange ($\rm{\Lambda}$), and 
multi-strange ($\rm{\Xi}$, $\rm{\Omega}$) in the most central \PbPb collisions is influenced by competing processes of strangeness enhancement and baryon-antibaryon annihilation.
Comparing these findings with measurements at LHC energies offers valuable insights into the underlying dynamics of the hadronic phase and their production dynamics.
\end{abstract}

\section{Introduction}
\label{sec:1}
\sloppy
High-energy heavy-ion (AA) collisions offer a unique opportunity to study the properties of the Quark-Gluon Plasma (QGP), a deconfined state of quarks and gluons, formed under extreme conditions of temperature and/or energy density \cite{Harris:1996zx}. As the dense medium created in these collisions evolves and cools, it eventually reaches a point where quarks and gluons recombine into hadrons, resulting in a hadron resonance gas (HRG). The study of the production yield of hadronic resonances offers valuable insight into the properties of the hadronic phase, which exists between chemical freeze-out (when inelastic collisions stop) and kinetic freeze-out (when elastic collisions stop)~\cite{ALICE:2022wpn}. 
The decay products of resonances can undergo regeneration or rescattering through elastic or pseudo-elastic interactions, which involve scattering via intermediate resonance states. These processes can modify the original yields of resonances produced before chemical freeze-out. 
In the rescattering process, if at least one decay product interacts elastically with other hadrons, the four-momentum information of the parent resonance is altered. This alteration prevents accurate reconstruction of the resonance, and thus the parent resonance particle is said to be lost. 
Conversely, regeneration occurs when pseudo-elastic interactions among hadrons recreate the parent resonance, leading to an increase in the reconstructed yield compared to the original resonance yield.
The dominance of rescattering or regeneration can be explored by analyzing the yield ratios of resonances to longer-lived hadrons with similar quark content as a function of system size. To investigate hadronic resonance production and their interactions, various resonances with different lifetimes, valence-quark flavors, masses, and spins have been studied at LHC~\cite{ALICE:2019etb, ALICE:2023egx, ALICE:2018qdv,ALICE:2019smg, ALICE:2018qdv, ALICE:2019xyr, ALICE:2022zuc} and RHIC~\cite{STAR:2006vhb, STAR:2004bgh, Gaudichet:2003jr} energies. These studies provide insight into the dynamics of the hadronic phase and the modification of resonance yields in different collision systems.

Recent measurements show that the yields of short-lived mesonic resonances, such as \rhmass (1.335 \fmc) and \kstarmass (4.16 \fmc), are significantly modified in the hadronic phase compared to longer-lived resonances like \phf meson (46.26 \fmc). This is attributed to the fact that short-lived resonances decay within the hadronic medium, where processes such as rescattering and regeneration affect their yields. In contrast, longer-lived resonances typically decay after the hadronic phase, when the medium has expanded sufficiently and the interactions are minimal, resulting in a little or no yield modification~\cite{ALICE:2023egx, ALICE:2019smg, ALICE:2019etb}.
For baryonic resonances, such as the \sigmass (5.0$-$5.5 \fmc) and \lambmass (12.54 \fmc), studies have shown the yield modification for \lambmass but not for \sigmass, despite the latter having a shorter lifetime in most central \PbPb collisions~\cite{ALICE:2018ewo}. Interestingly, such studies in small collision systems indicate that there is no significant yield modification for these resonances~\cite{ALICE:2019smg}. 

Recent experimental findings from high-multiplicity small system collisions show behavior similar to that observed in heavy-ion collisions, generating significant interest in such studies in these systems. The \kstarmass measurement shows a decreasing trend in yield with multiplicity, indicating the presence of a non-zero hadronic phase duration in high-multiplicity pp and \pPb collisions, as reported in Ref.~\cite{ALICE:2021ptz, ALICE:2021rpa}.

To study the dynamics of resonance production and their yield modification due to the hadronic phase, various approaches have been developed that
couple a hadronic cascade with or without fluid approximation for the dense medium formed in high energy collisions. EPOS with UrQMD combines hydrodynamic evolution with a hadronic afterburner, providing a hybrid framework that integrates both macroscopic and microscopic dynamics. In contrast, the AMPT model combines a partonic phase with a hadronic cascade based on the relativistic transport (ART) model to simulate final-state interactions, with the option to switch hadronic scattering effects on or off~\cite{Lin:2004en}. 
MUSIC with SMASH model (Simulating Many Accelerated Strongly-interacting Hadrons) is a stand-alone hadronic transport approach coupled to hydrodynamic evolution as an afterburner. The SMASH model accounts for hadronic cascade effects in a manner similar to  the UrQMD, but as a more recent framework it treats all interactions through explicit resonance formation and decay, employing PDG-consistent lifetimes, branching ratios, and conservation laws~\cite{Oliinychenko:2021enj}. In contrast, the UrQMD covers a broader energy range and remains a well-tested tool for hadronic cascades, having undergone decades of validation in several studies.
On the other hand, the hadron resonance gas (HRG) model with partial chemical equilibrium (PCE) is modeled using the law of mass action. In this framework, short-lived resonances can decay and regenerate, and their abundances remain in equilibrium with the particles produced in their decays~\cite{Motornenko:2019jha}.
Moreover, the findings from experimental measurements and theoretical studies indicate that the final resonance yields are influenced not only by their lifetimes but also by additional factors, such as the duration of the hadronic phase, the interaction cross-section of decay daughters, the freeze-out temperature, and the mean free path of the resonances.

\sloppy
In this study, we present results of various hadrons, including hadronic resonances, at midrapidity (\yrange~)  for pp collisions at \s = 13.6 \TeV and Pb--Pb collisions at \snn = 5.36 \TeV using the EPOS4 with UrQMD hadronic afterburner both ON and OFF. The terms ``with UrQMD" and ``without UrQMD" are used interchangeably to refer to simulations with UrQMD ON and UrQMD OFF, respectively, and may appear throughout the text.
The inclusion of UrQMD (Ultra-relativistic Quantum Molecular Dynamics) enables detailed modeling of the hadronic phase by simulating its effects on the system. These effects can significantly alter key observables such as transverse momentum (\pt) spectra, particle yields, and yield ratios, as well as collective phenomena such as flow and particle correlations. By comparing simulations with and without UrQMD to experimental data, the influence of the hadronic phase on resonance production has been explored.
The study also examines how strangeness enhancement and baryon-to-meson ratios influence the production of hadronic resonances and multistrange hadrons. This comparison provides valuable insights into the dynamics of the hadronic phase and its impact on final-state observables. The highest energies and finer high-multiplicity events in small systems are chosen to improve upon previous studies~\cite{Knospe:2021jgt, Knospe:2015nva} and to offer opportunities for future comparisons with experimental measurements.

\sloppy
Experimentally, hadronic resonances are reconstructed using the invariant-mass technique via the addition of 4-momenta of their decay daughters. The longer-lived decay daughters, such as charged pions, charged kaons, and (anti)protons, reach the detectors and are often identified through measurements of energy loss (\dEdx) in a Time Projection Chamber (\TPC) and/or velocity in a Time-of-Flight (\TOF) detector. The weakly decaying daughters, such as $\mathrm{K^{0}_{S}}$, $\rm{\Lambda}$, $\rm{\Xi}$ and $\rm{\Omega}$ can be selected based on their decay topologies, which puts further constraints. 
For this study, 5 million events were generated for pp collisions, both with and without UrQMD, while 1.5 million events were generated for \PbPb collisions under the same conditions.
The events are divided into various multiplicity classes based on the number of charged particles present within the pseudorapidity ranges $-3.7 < \eta < -1.7$ and $2.8 < \eta < 5.1$. This method follows an approach similar to that used by the ALICE experiment for pp and \PbPb collisions in the Run 2 studies.

Particles are selected from the generated data according to their unique EPOS IDs, with weak decays disabled to ensure that only primary particles contribute to the final yield calculations. After selecting resonances based on their unique EPOS IDs, they are flagged as either reconstructible or non-reconstructible. When a resonance decays, the EPOS4 model tracks its decay daughters. If any of the decay daughters undergo elastic interactions that alter their momenta, the parent resonance is flagged as non-reconstructible. As a result, the resonance is not included in the final yield measurements. This ensures that only resonances whose decay daughters have not undergone significant elastic scattering are counted, following a similar approach to experimental measurements, where resonance particles are reconstructed based on the momenta and trajectories of their decay products.
Table~\ref{table:1.1} lists the resonances and their properties, including decay channels, branching ratios, and lifetimes as in the UrQMD model. The decay channels included in Table \ref{table:1.1} are those commonly analyzed in experiments at LHC energies~\cite{ALICE:2014sbx}. The shorthand notation provided in the table will be used to represent the resonances throughout the text.
\begin{table}[htbp!]
\centering
\caption{Properties of selected resonances, including their decay channels, branching ratios, and lifetimes in the rest frame.}
\scalebox{0.89}{
\begin{tabular}{llllll}
\hline
\hline
 & & Decay & Branching & Lifetime \\
Resonance &  Shorthand & Channel & Ratio & (\fmc) \\
\hline
\vspace{0.12cm}
\(\rho(770)^{0}\) & $\rho^{0}$ & $\pi^{+} + \pi^{-} $ & 1.0 & 1.335 \\
\vspace{0.12cm}
\(K^{*}(892)^{0}\) & $\mathrm{K}^{*0}$ & $\pi^{-} + \mathrm{K}^{+}$ & 0.667 & 4.16 \\
\vspace{0.12cm}
\(\phi\)(1020) & $\phi$ & $\mathrm{K}^{+} + \mathrm{K}^{-}$ & 0.489 & 46.26 \\
\vspace{0.12cm}
\(\Delta^{++}\)(1232) & $\Delta^{++}$ & $\pi^{+} + \mathrm{p}$ & 1.0 & 1.69 \\
\vspace{0.12cm}
\(\Sigma(1385)^{+}\) & $\Sigma^{*+}$ & $\pi^{+} + \Lambda $ & 0.870 & 5.48 \\
\vspace{0.12cm}
\(\Sigma(1385)^{-}\) & $\Sigma^{*-}$ & $\pi^{-} + \Lambda $ & 0.870 & 5.01 \\
\vspace{0.12cm}
\(\Lambda\)(1520) & $\Lambda^{*}$ & $\mathrm{K}^{-} + \mathrm{p}$ & 0.225 & 12.54 \\
\(\Xi(1530)^{0}\) & $\Xi^{*0}$ & $\pi^{+} + \Xi^{-}$ & 0.667 & 22 \\
\hline
\hline
\end{tabular}
}
\label{table:1.1}
\end{table}
The paper is organized as follows. Section~\ref{sec:1} introduces hadronic resonances and their interactions within the hadronic gas. It also describes the event generation using the EPOS4 hydrodynamical model, along with particle selection methods and the criteria used for the estimation of multiplicity and centrality. Section~\ref{sec:2} discusses the EPOS4 model, outlining its key concepts and emphasizing the major advancements and differences compared to its predecessor, EPOS3. Section~\ref{sec:3} presents the results for various observables. Finally, Section~\ref{sec:4} summarizes the key findings of this study.

\section{EPOS4: Framework overview and key features}
\label{sec:2}
\sloppy
The EPOS4 model is a general purpose framework designed to investigate a wide range of observables in relativistic collisions, from proton-proton to nucleus-nucleus collisions, at energies ranging from several \GeV to multiple \TeV per nucleon~\cite{Werner:2023jps}. One of the key concepts underlying EPOS4 is parallel scattering, where multiple parton-parton scatterings occur simultaneously due to the extended reaction times in high-energy collisions. This is implemented using the Gribov-Regge (GR) theory~\cite{Gribov:1967vfb, Gribov:1968jf, Gribov:1972ri, Abramovsky:1973fm}, applying the S-matrix framework to manage multiple scatterings, which ensures the proper distribution of energy and momentum among the scatterings and the remnants of both the projectile and target. The EPOS4 model also accounts for those scenarios where the Dokshitzer-Gribov-Lipatov-Altarelli-Parisi (DGLAP) approach~\cite{Gribov:1972ri, Altarelli:1977zs, Dokshitzer:1977sg} becomes insufficient, particularly at small momentum fractions ($\rm{x} \ll 1$), where the linear parton evolution assumed by DGLAP breaks down due to the increasing density of partons. These small-$x$ regions, where parton densities are very high, require the inclusion of nonlinear effects arising from gluon-gluon fusion, leading to \enquote{saturation} phenomena~\cite{Gribov:1983ivg, McLerran:1993ka, McLerran:1993ni, Kovner:1995ts, Kovchegov:1996ty}. To address these saturation effects, the EPOS4 introduces a saturation scale that ensures proper factorization and binary scaling, which were violated in the previous EPOS3 model~\cite{Pierog:2013ria}. With this new approach, the model also offers a unique opportunity to distinguish between the primary and secondary scatterings: primary scatterings refer to instantaneous interactions involving initial nucleons and their partonic constituents at very high energies, whereas the secondary scatterings involve interactions among the products of string decays.

In this model, the particle production is based on the core-corona approach. At an early proper time $\tau_0$ of the collisions, prehadrons (which are intermediate states) are formed from pomerons or excited remnants~\cite{Werner:2023jps}. These prehadrons are categorized as either core or corona, depending on their likelihood to interact with the surrounding medium. This classification is based on their modified transverse momentum. The core prehadrons are situated in dense regions and form a thermalized medium that evolves hydrodynamically. In contrast, the corona prehadrons experience minimal interaction and hadronize directly.
The core undergoes hydrodynamic evolution with $\eta/s = 0.08$ and hadronizes on a hypersurface of constant energy density $\varepsilon_H = 0.57~\text{GeV/fm}^3$. In EPOS4, hadronization is not performed using the standard (grand-canonical) Cooper–Frye prescription, which assumes a smooth fluid-to-particle transition. Instead, the model employs a fully microcanonical hadronization approach that captures the abrupt hadronization in a rapidly expanding system at the critical energy density, followed by a random multi-hadron decay that maximizes entropy. This is suitable for both small and large systems from proton-proton to heavy-ion collisions. It provides a more realistic description of hadronization by enforcing exact quantum-number conservation per event. This improves baryon production, strangeness enhancement, and event-by-event fluctuations, especially in small systems, whereas the previous EPOS3 with GCE relied on statistical sampling, leading to artificial fluctuations in particle yields and correlations.
The model also improves on the core-corona procedure~\cite{Werner:2007bf, Werner:2010aa, Werner:2013tya}, dividing string segments based on energy loss and incorporating advanced methods to manage energy-momentum flow through the freeze-out hypersurface~\cite{Werner:2023jps}. These changes, along with a new microcanonical procedure~\cite{Werner:2023jps} to conserve energy-momentum and flavor throughout primary and secondary interactions, mark significant improvements over EPOS3, making EPOS4 a more robust tool for modeling high-energy collisions; detailed discussion can be found in Ref.~\cite{Werner:2023jps}.
Furthermore, to simulate the effects of the hadronic phase in high-energy collisions, the UrQMD model is employed as a hadronic afterburner. After hadronization, the produced hadrons are fed into the UrQMD, which is a microscopic transport model based on $2 \rightarrow n$ hadronic scatterings. The model uses experimentally measured reaction cross-sections for various resonances having different lifetimes, as its primary input. It includes baryon-baryon, meson-baryon, and meson-meson interactions, featuring approximately 60 baryonic species (along with their antiparticles) and around 40 mesonic states~\cite{Bleicher:1999xi}, thereby providing a comprehensive description to study hadronic interactions.
\section{Results and discussion}
\label{sec:3}
\sloppy
This section includes the \pt spectra, the \pt-differential and integrated yield ratios (including resonance-to-stable hadron, baryon-to-meson, and hadron yields to pion), and the \meanpt as a function of system size from pp to \PbPb collisions at LHC energies. Additionally, the time duration of the hadronic phase is estimated for short-lived resonances. The discussion of these observables aims to provide insight into the hadronic phase. Whenever available, the results are compared with the experimental data from the ALICE Collaboration.

\subsection{Transverse momentum \texorpdfstring{(\pt)}{Transverse momentum} spectra}
\label{sec:3.1}
\begin{figure}[!htbp]
\centering
 \includegraphics[width=0.48\textwidth]{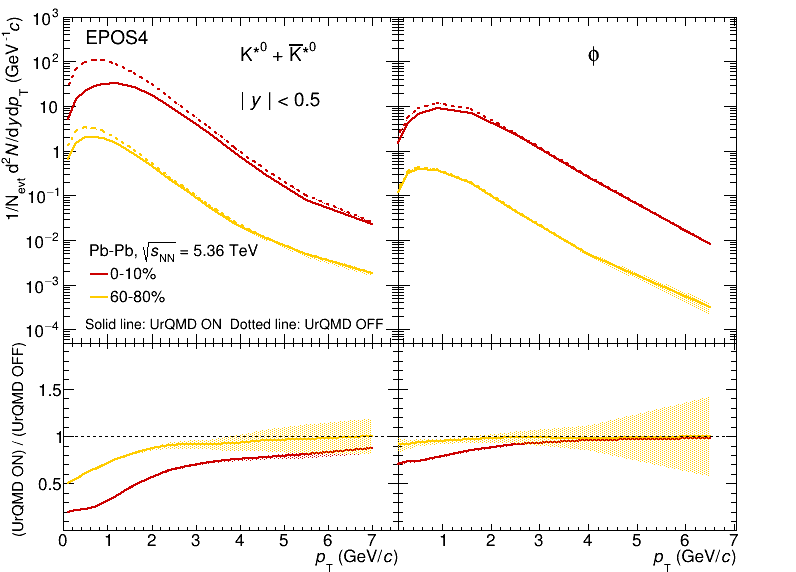}
\caption{\textbf{Upper panel:} The \pt spectra of \kstar and $\phi$ resonances in the midrapidity region for central (0--10\%) and peripheral (60--80\%) \PbPb collisions at \snn = 5.36 \TeV with EPOS4 in UrQMD ON and UrQMD OFF tunes. The solid lines represent measurements with UrQMD ON while dotted lines represent measurements with UrQMD OFF. \textbf{Lower panel:} The \pt-differential yield ratios of \kstar and $\phi$ for UrQMD ON to UrQMD OFF tunes. The bands represent the statistical uncertainties in the measurements.}
\label{fig:3.1}       
\end{figure}

\begin{figure}[!htbp]
\centering
 \includegraphics[width=0.48\textwidth]{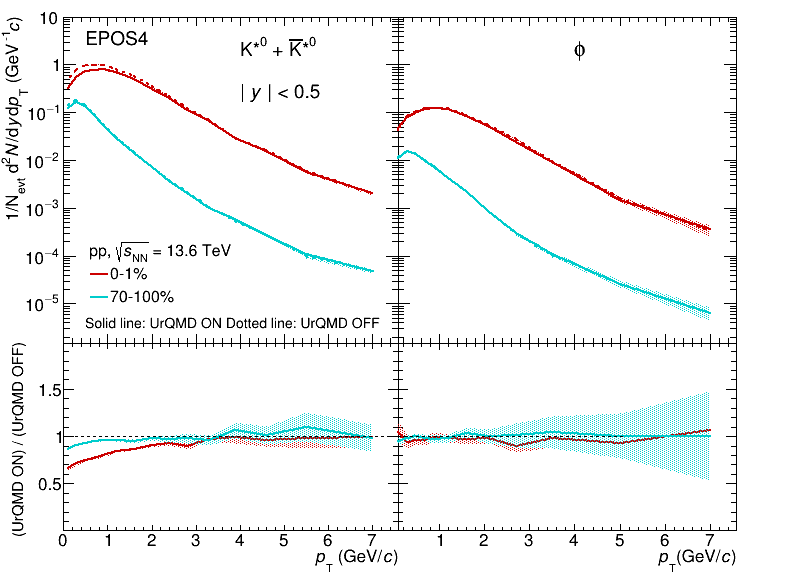}
\caption{\textbf{Upper panel:} The \pt spectra of \kstar and $\phi$ resonances in the midrapidity region for high multiplciity (0--1\%) and low multiplciity (70--100\%) in pp collisions at \s = 13.6 \TeV with EPOS4 in UrQMD ON and UrQMD OFF tunes. The solid lines represent measurements with UrQMD ON while dotted lines represent measurements with UrQMD OFF. \textbf{Lower panel:} The \pt-differential yield ratios of \kstar and $\phi$ for UrQMD ON to UrQMD OFF tunes. The bands represent the statistical uncertainties in the measurements.}
\label{fig:3.2}       
\end{figure}

\begin{figure}[!htbp]
\centering
 \includegraphics[width=0.48\textwidth]{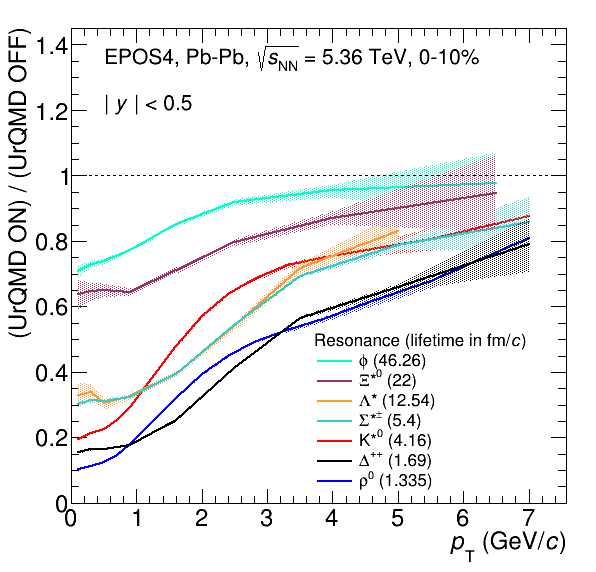}
\caption{The comparison of \pt-differential spectra of hadronic resonances with UrQMD to the spectra without UrQMD in central (0-10\%) \PbPb collisions with EPOS4. The bands in the measurements represent the statistical uncertainty.}
\label{fig:3.3}       
\end{figure}

\begin{figure*}[!htbp]
\centering
 \includegraphics[width=0.75\textwidth]{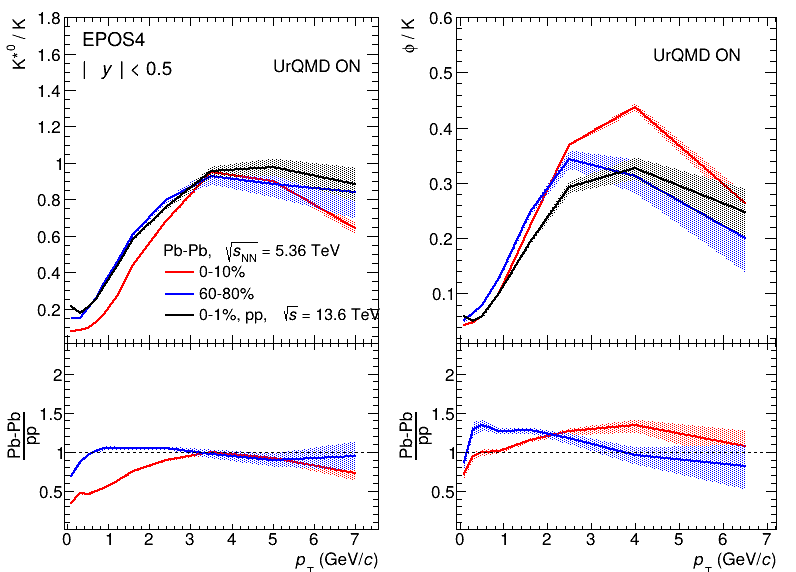}
\caption{\textbf{Upper panel:} The \pt-differential ratios $\rm{K^{*0}/K}$ (left) and $\rm{\phi/K}$ (right) in the central (0--10\%) and peripheral (60--80\%) \PbPb collisions at \snn = 5.36 \TeV and also in the high multiplicity (0-1\%) pp collisions at \s = 13.6 \TeV with UrQMD. \textbf{Lower panel:} The comparison of \pt-differential ratios in the central and peripheral \PbPb collisions to the high multiplicity pp collisions. The bands in the measurements represent statistical uncertainty.}
\label{fig:3.4}       
\end{figure*}

\begin{figure*}[!htbp]
\centering
 \includegraphics[width=0.75\textwidth]{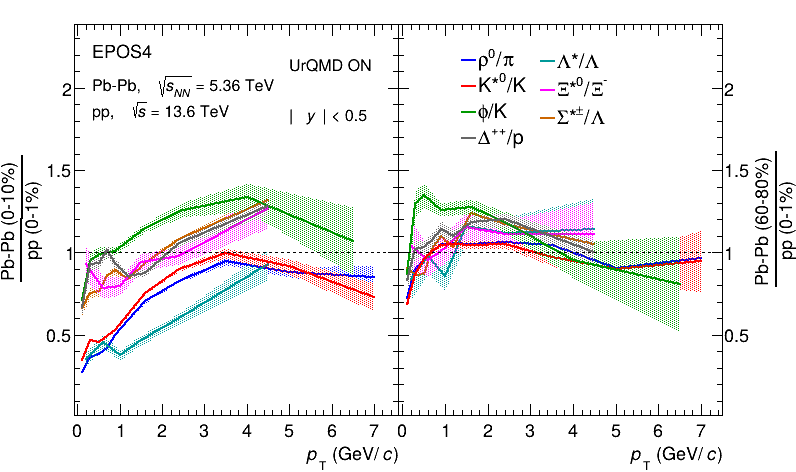}
\caption{The comparison of \pt-differential ratios $\rho^{0}/\pi$, $\rm{K^{*0}/K}$, $\rm{\phi/K}$, $\rm{\Delta^{++}/p}$, $\rm{\Lambda^{*}/\Lambda}$, $\rm{\Xi^{*0}/\Xi^{-}}$ and $\rm{\Sigma^{*\pm}/\Lambda}$ in the central (0--10\%, left) and peripheral (60--80\%, right) \PbPb collisions at \snn = 5.36 \TeV to the high multiplicity (0-1\%) pp collisions at \s = 13.6 \TeV with UrQMD. The bands in the measurements represent statistical uncertainty.}
\label{fig:3.5}     
\end{figure*}

\sloppy
To investigate the impact of the hadronic phase on resonance production, analyzing the shape of the \pt spectra can provide details about their production dynamics and interactions. The \kstar and \phim mesons are ideal candidates for such studies due to their similar masses but significantly different lifetimes: \kstar has a relatively short lifetime of approximately 4.16 \fmc, while the \phim meson is long-lived, with a lifetime of about 46.2 \fmc, differing by a factor of 10.
The upper panels of Fig.~\ref{fig:3.1} show the \pt spectra of \kstar (left) and \phim (right) from EPOS4 with UrQMD ON (solid lines) and OFF (dotted lines) for the centrality classes 0--10\% and 60--80\% at midrapidity (\yrange~) in \PbPb collisions at \snn = 5.36 \TeV. The lower panels represent the ratios of \pt spectra with UrQMD ON to those with UrQMD OFF. A significant difference is observed between UrQMD ON and OFF at low \pt for \kstar. The effect is more pronounced for central (0--10\%, red lines) collisions than peripheral (60--80\%, yellow lines) collisions. Similarly, the same ratios for pp collisions in the multiplicity class 0--1\% (red) and 70--100\% (cyan) at \s = 13.6 \TeV are shown in Fig.~\ref{fig:3.2}.
A similar  feature is also observed at low \pt in high-multiplicity (0--1\%) pp collisions when comparing \kstar to \phim. Moreover, the magnitude of this effect in high-multiplicity pp collisions is comparable to that observed in peripheral \PbPb collisions. The average charged-particle multiplicity ($\langle\mathrm{d}N_\mathrm{ch} / \mathrm{d}\eta\rangle_{\eta<0.5}$) in 0--1\% high-multiplicity pp collisions ($\simeq$ 27) approaches that observed in 60--100\% peripheral \PbPb collisions ($\simeq$ 36). As similar $\langle\mathrm{d}N_\mathrm{ch}/\mathrm{d}\eta\rangle_{\eta<0.5}$ values are achieved across different collision systems, the evolution of resonance yields can be primarily attributed to event multiplicity rather than colliding systems and energies.
In the EPOS4 simulations with UrQMD, the suppression of \kstar yields is evident when compared to results without UrQMD and to the \phim production. This suppression is attributed to the short lifetime of \kstar, as its decay daughters undergo rescattering in the hadronic phase, preventing reconstruction of the parent resonance and thereby reducing its observed yield compared to the original yield of the resonance. The effect is particularly pronounced at low \pt in central \PbPb collisions and high-multiplicity pp collisions, where a large and dense medium is formed and the probability of rescattering is high. 
At high \pt and in low-multiplicity events, the ratios approach unity because the high-\pt particles escape the hadronic medium more rapidly and the likelihood of rescattering effect is negligible. In comparison, the effect is less pronounced for \phim, as expected due to its longer life. The EPOS4 results with UrQMD qualitatively support the observations reported by the ALICE Collaboration at LHC energies~\cite{ALICE:2021ptz}, demonstrating that the EPOS4 model with UrQMD effectively captures the impact of the hadronic phase on resonance production. In particular, the rescattering of decay daughters plays a dominant role at low \pt for short-lived resonances.
\sloppy
To further investigate the effects of the hadronic phase, the ratios of $p_\mathrm{T}$-differential yields with UrQMD ON to those with UrQMD OFF are studied for various hadronic resonances with lifetimes ranging from 1 to 47~fm/$c$. The results for central (0--10\%) \PbPb collisions at $\sqrt{s_{\mathrm{NN}}} = 5.36$~TeV are shown in Fig.~\ref{fig:3.3}. A clear trend is observed in the suppression pattern, ordered as: $\rho^0 < \Delta^{++} < \mathrm{K}^{*0} < \Sigma^{*\pm} \sim \Lambda^{*} < \Xi^{*0} < \phi$. The suppression becomes more pronounced as the resonance lifetime decreases, assuming negligible regeneration effects. However, 
the $\Sigma^{*\pm}$ shows a suppression similar to that of $\Lambda^{*}$ despite having a significantly shorter lifetime. This can be attributed to the contribution from regeneration, which is more effective for the $\Sigma^{*\pm}$ than for the $\Lambda^{*}$. Since regeneration counteracts the effects of rescattering, it partially offsets the suppression. As a result, even though the $\Sigma^{*\pm}$ has a lifetime nearly half that of the $\Lambda^{*}$, the net suppression appears comparable.
These observations show that the final yields of resonances depend not only on their lifetimes but also on additional factors such as decay daughters cross section, chemical freeze-out conditions, and the lifetime of the hadronic phase. In particular, for baryonic resonances, the interaction cross sections of their decay daughters play a crucial role in modifying the yields during the hadronic phase.
\sloppy
The particle ratios are used as key tools for distinguishing various processes such as rescattering, regeneration, strangeness enhancement, and baryon-to-meson production mechanisms, which give information about the production and interactions of hadrons in the hadronic medium produced in high-energy collisions.
Specifically: (a) resonance-to-stable hadron yield ratios provide insight into the \pt distributions of hadrons with similar quark content but differing masses, and (b) baryon-to-meson resonance yield ratios enable the comparison of hadrons with similar masses but differing baryon numbers and quark compositions. 
(c) strangeness enhancement through relative yield ratios of hadrons to pions.
To quantify the $p_\mathrm{T}$-dependence of the rescattering effect, the upper panels of Fig.~\ref{fig:3.4} present a comparison of the $p_\mathrm{T}$-differential yield ratios $\mathrm{K}^{*0}/\mathrm{K}$ (left) and $\phi/\mathrm{K}$ (right) in central (0--10\%) and peripheral (60--80\%) \PbPb collisions at $\sqrt{s_{\mathrm{NN}}} = 5.36$~TeV, as well as in high-multiplicity (0--1\%) pp collisions at $\sqrt{s} = 13.6$~TeV, all with the UrQMD hadronic afterburner ON. The lower panels display the double ratios, where the \PbPb results are divided by those from high-multiplicity pp collisions. 
Although the center-of-mass energies differ between the pp and \PbPb systems, earlier observations as mentioned in Fig.~\ref{fig:3.1} and ALICE measurements~\cite{ALICE:2021ptz, ALICE:2018ewo} have shown that resonance yields evolve smoothly with charged-particle multiplicity rather than with the collision system or energy. Therefore, the 0--1\% high-multiplicity pp collisions serve as an appropriate baseline for comparison, particularly with peripheral \PbPb collisions, providing further insights into the role of the hadronic phase in describing the resonance yield. For $\mathrm{K^{*0}/K}$, the double ratio is found to be suppressed at low $p_\mathrm{T}$, particularly in central \PbPb collisions compared to peripheral ones. This suppression is primarily driven by rescattering effects in the hadronic phase, which affect the $\mathrm{K^{*0}}$ yield. The use of the $\mathrm{K^{*0}/K}$ ratio effectively cancels out any strangeness-related effects, isolating the impact of rescattering. In contrast, the $\mathrm{\phi/K}$ double ratio does not exhibit significant suppression or centrality dependence, consistent with the longer lifetime of the $\phi$ meson and its reduced sensitivity to hadronic phase interactions.
Peripheral \PbPb collisions and high-multiplicity pp collisions exhibit similar \pt-differential yield ratios. The observed similarities demonstrate that the production of resonances is mainly governed by charged-particle multiplicity, irrespective of the colliding system. Similar behavior has also been reported by the ALICE experiment, and the model results support the experimental measurements.
Similarly, \pt-differential yield ratios for various hadronic resonances are shown in Fig.~\ref{fig:3.5} for centrality 0--10\% (left) and peripheral 60--80\% (right) in \PbPb collisions at \snn = 5.36 TeV. At low \pt, these ratios exhibit a specific ordering: $\lstar/\Lambda$ $<$ $\rho^{0}/\pi$ $<$ $\rm{K^{*0}/K}$ $<$ $\rm{\Xi^{*0}/\Xi^{-}}$  $\approx$ $\rm{\Sigma^{*\pm}/\Lambda}$ $<$ \delstar/p $<$ $\rm{\phi/K}$. 
Despite the longer lifetime of the $\Lambda^{*}$, it exhibits a larger suppression compared to other short-lived resonances, a trend also observed in the \pt-integrated yield ratios for the most central \PbPb collisions (discussed later in Section~\ref{sec:3.4}). This can be attributed to its relatively low regeneration probability in the hadronic phase. Although short-lived resonances such as $\mathrm{K}^{*0}$ and $\Sigma^{*\pm}$ are more affected by rescattering, their suppression is partially compensated for by stronger regeneration due to larger hadronic cross sections with their decay daughters. In contrast, $\Lambda^{*}$, although less prone to rescattering due to its longer lifetime, experiences less regeneration, leading to a stronger overall suppression. EPOS4 with UrQMD supports this interpretation, showing that the regeneration contributions follow the order: $R_{\rm{K+p}} < R_{\rm{K}+\pi} <  R_{\Lambda+\pi}$.
\subsection{Baryon-to-meson ratios}
\label{sec:3.2}
\begin{figure}[!htbp]
\centering
 \includegraphics[width=0.48\textwidth]{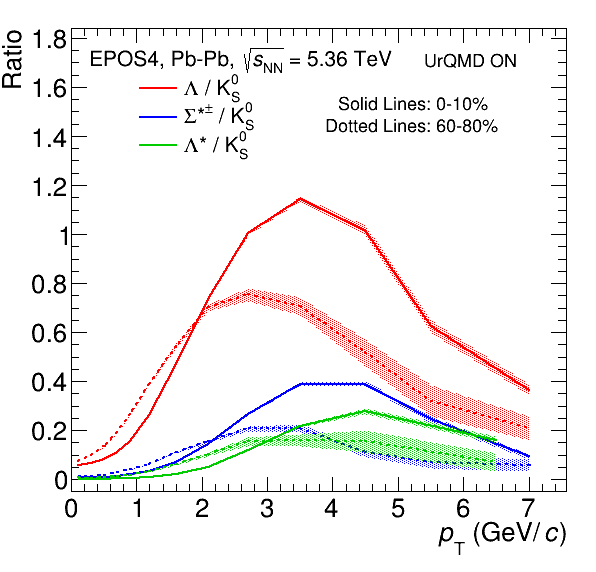}
\caption{The \pt-differential ratios $\rm{\Lambda/K^{0}_{S}}$, $\rm{\Sigma^{*\pm}/K^{0}_{S}}$, and $\rm{\Lambda^{*}/K^{0}_{S}}$ in central (0–10\%, solid lines) and peripheral (60–80\%, dotted lines) \PbPb collisions at \snn = 5.36 \TeV, simulated using EPOS4 with UrQMD. The shaded bands represent statistical uncertainties in the measurements.}
\label{fig:3.6}       
\end{figure}

\sloppy
Particle ratios, particularly baryon-to-meson ratios, are essential
for understanding mass-dependent radial flow and the production mechanisms of baryons and mesons. The baryon-to-meson ratios of several hadrons, including resonances such as \kstar, \phim, \lstar, and \sigmstar, are discussed using EPOS4 calculations with and without UrQMD. The baryon-to-meson ratios ($\rm{\Lambda}$, \lstar, and \sigmstar  to \kzero) for most central (0-10\%) and peripheral (60-80\%) \PbPb collisions using EPOS4 with UrQMD are shown in Fig. \ref{fig:3.6}. A clear mass-dependent enhancement ( \lstar/\kzero $<$ \sigmstar/\kzero $<$ $\rm{\Lambda}$/\kzero) is observed at the intermediate \pt.  The enhancement in the ratios decreases and the peak position shifts toward the higher \pt region with increasing baryon masses. 
Similar behavior is also observed in the ratios p/$\pi$  and $\lmb/\kzero$  for central \PbPb collisions, consistent with observations reported by the ALICE Collaboration in Ref.~\cite{ALICE:2019hno, ALICE:2020jsh}. The effect of radial flow is evident in Fig.~\ref{fig:3.6} for the most central collisions, where a clear mass-dependent trend is observed. Radial flow boosts low-momentum particles to intermediate $p_{\mathrm{T}}$, with a more pronounced effect for heavier particles.

\begin{figure}[!htbp]
\centering
 \includegraphics[width=0.44\textwidth]{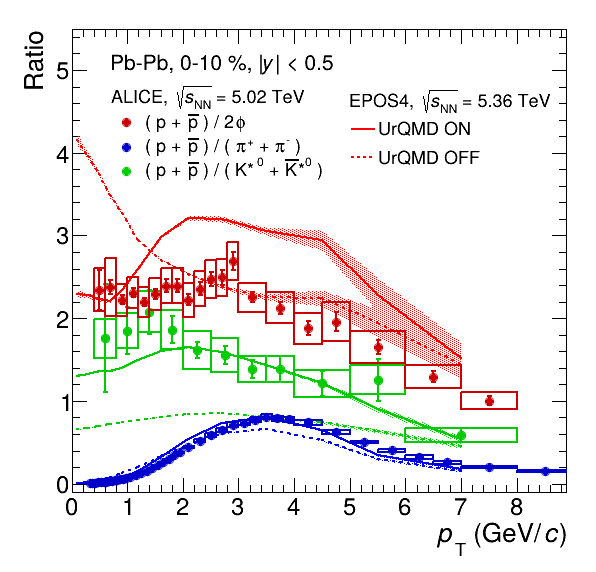}
\caption{The \pt-differential ratios p/$\phi$, p/$\pi$ and p/\kstar in the central \PbPb collisions with ALICE at \snn = 5.02 \TeV (markers) and using EPOS4 model with UrQMD (solid lines) and without UrQMD (dotted lines)  at \snn = 5.36 \TeV. The bars and boxes in the ALICE measurements represent statistical and systematic uncertainties, respectively. The shaded area in the model calculations represent statistical uncertainty. }
\label{fig:3.7}       
\end{figure}

To further investigate whether the enhancement at intermediate $p_{\mathrm{T}}$ arises primarily from radial flow and hadronization picture, the  particles with similar masses (p, $\mathrm{K}^{*}$, and $\phi$) are studied. This choice helps minimize the influence of mass-dependent effects associated with radial flow. The ratios p/$\phi$ and p/$\mathrm{K}^{*}$ are calculated using EPOS4 with and without UrQMD for 0--10\% central \PbPb\ collisions and compared with ALICE measurements, as shown in Fig.~\ref{fig:3.7}. The EPOS4 model with UrQMD shows a relatively flat behavior in the p/$\phi$ ratio up to $p_{\mathrm{T}} < 1.5$~GeV/$c$, but overestimates the ALICE data at higher $p_{\mathrm{T}}$. In contrast, the ALICE data exhibit a nearly flat p/$\phi$ ratio up to $p_{\mathrm{T}} \sim 4$~GeV/$c$. Although the model qualitatively captures the trend, further improvements are needed for a more quantitative agreement with the experimental data. The \(\mathrm{p}/\mathrm{K}^{*0}\) ratio measured by ALICE exhibits a slight decreasing trend, which is consistently reproduced by the model with UrQMD hadronic afterburner. This behavior is expected, as the \(\mathrm{K}^{*0}\) spectral shape is strongly modified by hadronic phase effects, particularly in the low-\(p_{\mathrm{T}}\) region. Furthermore, no significant enhancement is observed in the \(\mathrm{p}/\mathrm{K}^{*0}\) ratio at intermediate \(p_{\mathrm{T}}\). In contrast, the \(\mathrm{p}/\pi\) ratio shows a clear enhancement at intermediate \(p_{\mathrm{T}}\), as observed measurement by the ALICE Collaboration. This enhancement is primarily due to radial flow, which preferentially boosts protons to higher momentum compared to pions because of their larger mass at a common velocity. In the EPOS4 model, the core of the system evolves hydrodynamically, followed by MCE hadronization that converts the fluid into particle degrees of freedom to form final-state hadrons. The subsequent hadronization, combined with collective flow, imparts a mass-dependent momentum boost ($p = m v$, where $m$ and $p$ are the hadron mass and momentum, and $v$ is the collective flow velocity). This boost shifts heavier hadrons to higher momentum compared to lighter ones, reproducing the mass-dependent behavior in the $p_{\mathrm{T}}$-spectra, particularly in the most central collisions. In addition to collective flow, differences in production mechanisms for baryons and mesons also contribute to the enhancement of these ratios. This is particularly evident in peripheral collisions, where strong radial flow is not expected, yet an enhancement at intermediate $p_{\mathrm{T}}$ is still observed. It arises from the core–corona picture and the exact conservation handled by MCE hadronization~\cite{Werner:2023jps}, leading to species-dependent behavior. These effects provide a possible explanation for baryon-to-meson enhancement in the EPOS4, complementary to quark coalescence mechanisms~\cite{Fries:2003vb}.
At high \pt, both experimental data and EPOS4 predictions show a decreasing trend across all particle species. In EPOS4, particles at high \pt are predominantly produced from the corona, where fragmentation rather than collective hydrodynamics governs the production. Consequently, the baryon-to-meson ratios decrease with \pt, reflecting the dominance of a common fragmentation mechanism in the production of both baryons and mesons at high \pt~\cite{Werner:2023jps}.
Furthermore, the EPOS4 model without UrQMD fails to describe the \( \mathrm{p}/\mathrm{K}^{*0} \) ratio except at high \pt, and also does not reproduce the observed enhancement in the \( \mathrm{p}/\pi \) ratio at intermediate \pt. Thus, incorporating the UrQMD hadronic afterburner is essential in EPOS4 to accurately describe the measured baryon-to-meson ratios. This shows that although protons, \kstar, \phim have different quark content, the mass of hadrons and the hadronic phase effects on short-lived resonances play an important role in the determination of the spectral shapes of these particles.
From the above discussion, it is clear that radial flow dominates over the hadronization mechanism in driving the enhancement of baryon-to-meson ratios in the intermediate$p_{\mathrm{T}}$ region for the most central collisions. When one of the hadrons is a resonance, require UrQMD to properly account for additional hadronic phase effects to describe these ratios.
\subsection{Average transverse momentum \texorpdfstring{(\meanpt)}{mean transverse momentum}}
\label{sec:3.3}
\begin{figure*}[!htbp]
\centering
 \includegraphics[width=0.8\textwidth]{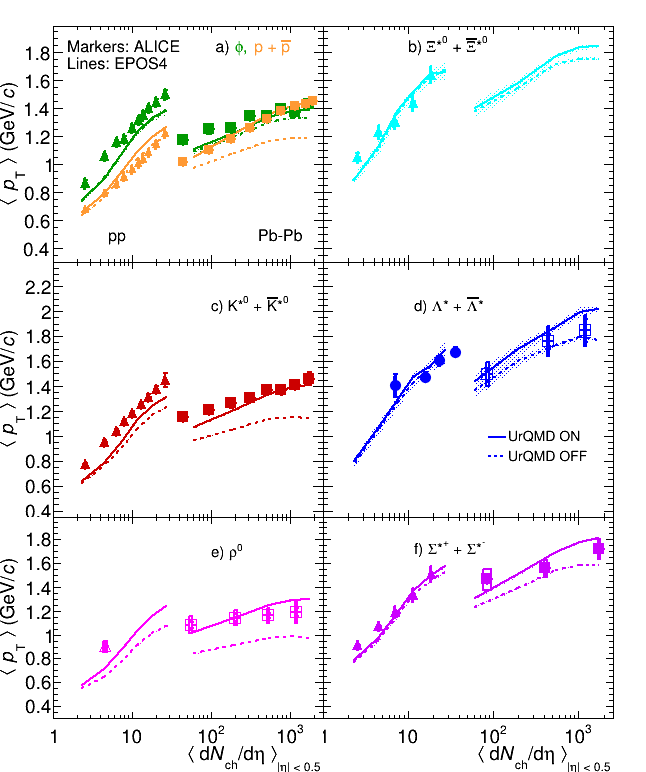}
\caption{Average transverse momentum of protons, mesonic resonances ($\rho^{0}$, \kstar, and $\phi$) and baryonic resonances ($\rm{\Sigma^{*\pm}}$, $\rm{\Lambda^{*}}$ $\rm{\Xi^{*0}}$) as a function of charged particle multiplicity density at intermediate-rapidity. Markers indicate ALICE measurements in pp system at \s = 2.76 \TeV (open triangles)~\cite{ALICE:2018qdv} and \s = 13 \TeV (solid triangles)~\cite{ALICE:2019etb, ALICE:2023egx, ALICE:2020nkc}, in \pPb system at \snn = 5.02 \TeV (soild circles)~\cite{ALICE:2019smg}, and in \PbPb system at \snn = 2.76 \TeV (open squares)~\cite{ALICE:2018ewo, ALICE:2018qdv} and \snn = 5.02 TeV (soild squares) ~\cite{ALICE:2021ptz, ALICE:2022zuc, ALICE:2019hno}. The lines represent EPOS4 predictions for pp collisions at \s = 13.6 \TeV and for \PbPb collisions at \snn =5.36 \TeV with UrQMD (solid lines) and without UrQMD (dotted lines). The statistical and systematic uncertainties in the ALICE data are represented by bars and boxes, respectively while the statistical uncertainty in the model measurements are represented by bands.}
\label{fig:3.8}       
\end{figure*}
\begin{figure*}[!htbp]
\centering
 \includegraphics[width=0.8\textwidth]{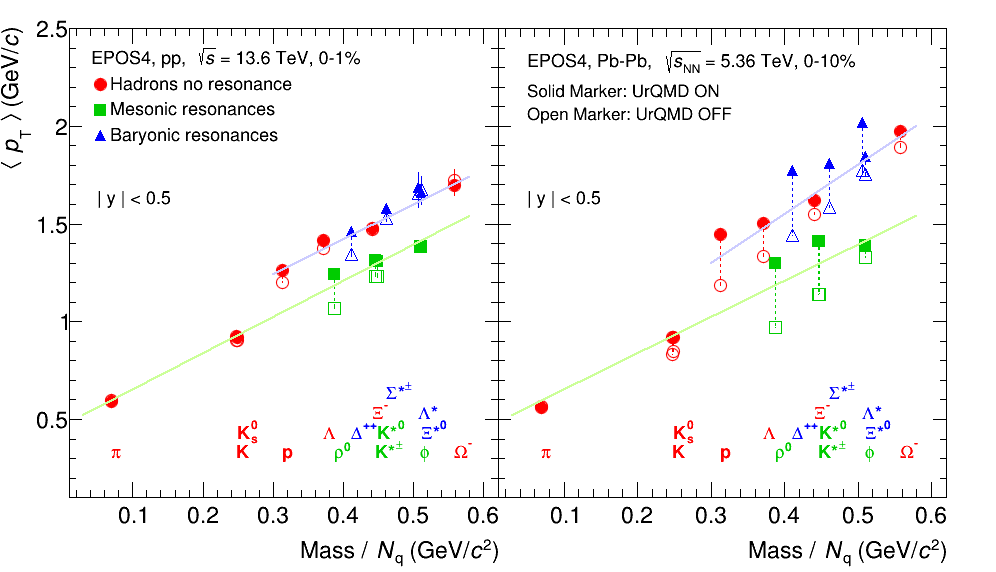}
\caption{The average transverse momentum of hadrons in the midrapidity region plotted as a function of the hadron mass, scaled by the number of valence quarks, in high multiplicity (0-1\%) pp collisions at \s = 13.6 \TeV (left panel) and in central (0-10\%) \PbPb collisions at \snn = 5.36 \TeV (right panel) using the EPOS4 model with UrQMD (solid markers) and without UrQMD (open markers). The green and blue lines represent linear functions. The statistical uncertainty in the measurements is represented by bars.}
\label{fig:3.9}       
\end{figure*}
The \meanpt of various hadronic resonances  as a function of average charged particle multiplicity  is shown in Fig.~\ref{fig:3.8} for midrapidty in pp collisions at \s = 13.6 \TeV and in \PbPb collisions at \snn = 5.36 \TeV. 
These calculations are obtained using EPOS4 with and without UrQMD hadronic afterburner. The results are also compared with the measurements from ALICE Collaboration  for pp collisions at \s = 2.76 and 13 \TeV~\cite{ALICE:2019etb, ALICE:2018qdv, ALICE:2023egx} and \PbPb collisions at \snn =  2.76 and 5.02 \TeV~\cite{ALICE:2016sak, ALICE:2019etb, ALICE:2022zuc}, when available.
The \meanpt increases with both charged particle multiplicity and the mass of the hadron species. The steeper increase of \meanpt with charged particle multiplicity observed in pp collisions compared to \(\mathrm{Pb\mbox{-}Pb}\) collisions is well reproduced by the EPOS4 model and consistent with ALICE measurements. This trend reflects the distinct underlying dynamics governing small and large systems and can be attributed to the interplay of several mechanisms. In particular, treatment of the core-corona picture, where a core evolves hydrodynamical with strong collective flow  coexists with a non-flowing corona, microcanonical hadronization of the core, exact conservation and possible parton saturation phenomena contribute to the observed behavior. These factors result in a sharper rise of \meanpt in pp events, especially at high multiplicities, while in \PbPb collisions, leads to a smoother evolution of \meanpt. A more detailed discussion of the non-monotonic behavior of \meanpt as a function of multiplicity can be found in Ref.~\cite{Werner:2023jps}.

The EPOS4 model generally reproduces the observed \meanpt trends in \PbPb collisions, demonstrating good agreement with experimental data across various centrality classes. In contrast, for pp collisions, EPOS4 tends to slightly underestimate the \meanpt values, particularly at higher multiplicities. The \meanpt values for short-lived resonances such as \rh, \kstar show significant changes between with and without UrQMD for \PbPb collision, whereas the effect is less pronounced for pp collisions. For the $\phi$-meson, the difference in \meanpt between the two cases (with and without UrQMD) is smaller because its decay daughters interact less with the hadronic medium due to its longer lifetime.
The change in \meanpt can be attributed to modifications in the spectral shape, as shown in Fig.~\ref{fig:3.1} and Fig.~\ref{fig:3.2}. For short-lived resonances, their decay daughters undergo substantial hadronic interactions in the hadronic phase, leading to pronounced modifications. This effect is strongest in central \PbPb collisions and decreases with lower multiplicity, consistent with the shorter hadronic phase lifetimes discussed later in Section~\ref{sec:3.5}. 
For baryonic resonances such as $\rm{\Sigma^{*\pm}}$, $\rm{\Lambda^{*}}$, and $\rm{\Xi^{*0}}$, no significant change in \meanpt is observed for small systems (up to $\langle\mathrm{d}N_\mathrm{ch}/\mathrm{d}\eta\rangle_{|\eta|<0.5}$ $\sim$ 27). However, in \PbPb collisions, both the spectral shape and the \meanpt values exhibit significant modifications due to the extended hadronic phase.
Particles with similar masses, such as the proton and the $\phi$-meson, show comparable \meanpt values in central \PbPb collisions, consistent with observation reported by ALICE measurements~\cite{ALICE:2021ptz, ALICE:2019hno}. This behavior is consistent with expectations from hydrodynamical models, which predict that $\langle p_{\mathrm{T}} \rangle$ depends mainly on particle mass in such collisions. The EPOS4 model calculations also support this trend. The \(\langle p_{\mathrm{T}} \rangle\) value of protons show a slight difference between  with and without UrQMD. This indicates that feed-down contributions from higher resonance decay daughters in the hadronic phase also modified the final proton \(p_{\mathrm{T}}\) spectrum. It is important to properly handel feed-down contribution on experimental measurement. The $\langle p_{\mathrm{T}} \rangle$ of protons shows only a slight difference with and without UrQMD. This indicates that decay daughters from higher-mass resonances in the hadronic phase also modify the final proton \pt spectrum. Properly accounting for these feed-down contributions from higher-mass resonances is important in experimental measurements.
In contrast, the \pt spectra of the \(\phi\)-meson is minimally affected by the hadronic phase, owing to its relatively long lifetime and small interaction cross-section with the hadronic medium (In Fig.~\ref{fig:3.7}, the \(p_{\mathrm{T}}\)-differential \(\mathrm{p}/\phi\) ratio from EPOS4 exhibits a flat behavior at low \(p_{\mathrm{T}}\), consistent with the data, while it overestimates the measurements for \(p_{\mathrm{T}} > 1.5~\mathrm{GeV}/c\)). The turn on of the hadronic cascade (UrQMD ON) generally increases $\langle p_{\mathrm{T}} \rangle$, bringing the prediction of the model closer to the experimental measurements. This effect demonstrates the crucial role of hadronic rescattering in modifying the $p_{\mathrm{T}}$ spectra during the late stages of the collision. 

The \meanpt as a function of  reduced mass (Mass/{$N_\mathrm{q}$}), where the mass of hadron is scaled by the number of valence quarks, is calculated for various light-flavor hadrons, including resonances for high multiplicity (0-1\%) pp collisions at \s = 13.6 \TeV and central (0-10\%) \PbPb collisions at \snn = 5.36 \TeV using the EPOS4 model, both with and without UrQMD, as shown in Fig.~\ref{fig:3.9}. The \meanpt follows a linear trend with increasing mass. Stable mesons (red markers) and mesonic resonances (green markers) are observed to generally follow a linear trend, aligning closely with the linear fit function represented by the green line for both UrQMD ON and OFF. This behavior is expected for a system that has  a common transverse expansion velocity ($\langle \beta_{\mathrm{T}} \rangle$), where the mean transverse momentum of the hadrons is  approximately proportional to their mass,
$\langle p_{\mathrm{T}} \rangle \approx m \times \langle \beta_{\mathrm{T}} \rangle$, resulting in a linear increase of $\langle p_{\mathrm{T}} \rangle$ with particle mass. Similar observations were observed in the experimental measurements, as discussed in Ref.~\cite{ALICE:2020jsh}. 

The observation that baryons and mesons group separately in \meanpt after scaling the hadron mass with their constituent quark numbers, with roughly parallel trends rather than collapsing onto a single curve, shows that collectivity is not fully carried over from the partonic stage to the hadronic final state and that hadronization dynamics affect baryons and mesons differently. It also indicates that hadronization is not purely coalescence-driven, even though \meanpt is dominated mainly by low \pt regions, where radial flow is the major contributor. Since EPOS4 does not explicitly implement a coalescence picture, the different grouping of baryons and mesons arises from hadronization mechanisms, including the core–corona picture, MCE hadronization, and hadronic rescattering, all of which introduce species-dependent effects. The ideal grouping of baryons and mesons can be understood through purely recombination-based hadronization models, and testing similar scaling with other observables, such as azimuthal anisotropy flow coefficients ($v_{n}$), could provide further insight into the observed differences, which is beyond the current scope of this study.

However, the resonances \rh, \kstar, \kstarpm, and \delstar slightly deviate from the linear trend, with different values observed between UrQMD ON and OFF. This deviation is expected due to the finite hadronic phase in high-multiplicity (0--1\%) pp collisions, which leads to the modification of the spectral shape of short-lived resonances.
Similarly, the observed deviation from the linear function increases for \PbPb collisions compared to pp collisions. Hadronic resonances such as \rh, \kstar, \kstarpm, \delstar, \sigm, and \lstar show significant deviations from the expected linear trend due to the larger size of the system. The value of \meanpt is higher for short-lived resonances when UrQMD is ON compared to when it is OFF for most central \PbPb collisions. This shows that the rescattering effect leads to an increased value of the \meanpt.
Additionally, for stable hadrons such as the proton and \lmb, the \meanpt values deviate from the linear trend and differ between the cases with UrQMD ON and OFF. This indicates that the spectral shape and \meanpt of protons and \lmb are modified in the hadronic phase due to contributions from the decay of higher resonance states, since they are likely decay daughters.
\subsection{Resonance to non-resonance yield ratios }
\label{sec:3.4}
\begin{figure*}[!htbp]
\centering
 \includegraphics[width=1.0\textwidth]{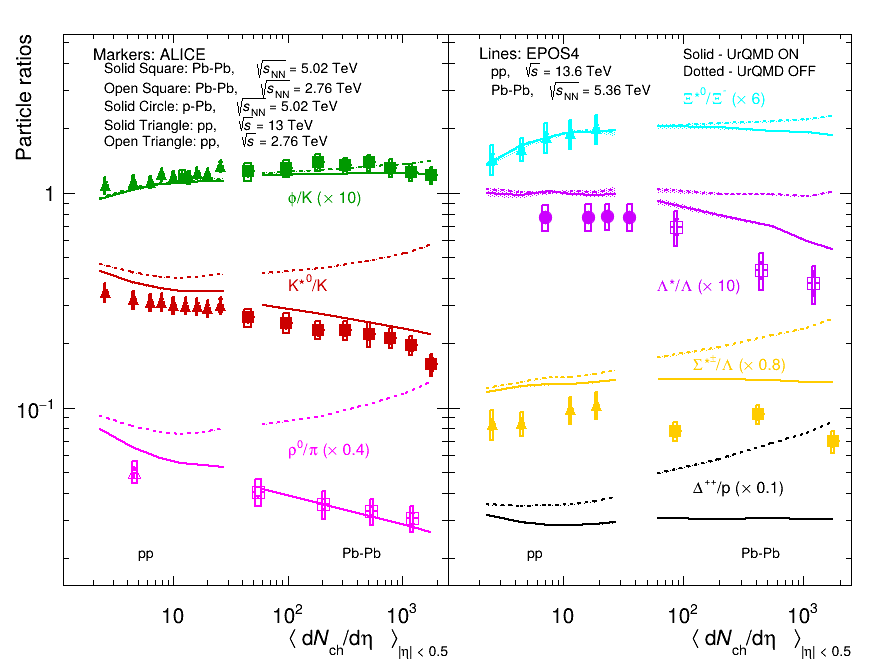}
\caption{Left panel shows the ratios of mesonic resonances to stable mesons yield while the right panel shows the ratios of baryonic resonances to stable baryons yield. Different markers represent ALICE results in pp collisions at \snn = 13 \TeV (solid triangles)~\cite{ALICE:2019etb, ALICE:2023egx} and \snn = 2.76 \TeV (open triangles)~\cite{ALICE:2018qdv}, \pPb collisions at \snn = 5.02 \TeV (solid circles)~\cite{ALICE:2019smg}, \PbPb collisions at \snn = 2.76 \TeV (open squares)~\cite{ALICE:2018qdv} and \snn = 5.02 \TeV (solid squares)~\cite{ALICE:2019xyr, ALICE:2022zuc}. The statistical and systematic uncertainties in the data are represented by bars and boxes, respectively. The lines represent predictions from EPOS4 in pp collisions at \snn = 13.6 \TeV and in \PbPb collisions at \snn = 5.36 \TeV with UrQMD (solid) and without UrQMD (dotted). The shaded area indicates the statistical uncertainty in the model calculations.}
\label{fig:3.11}       
\end{figure*}
 The resonance-to-stable-hadron yield ratios as a function of charged-particle multiplicity ($\langle\mathrm{d}N_\mathrm{ch}/\mathrm{d}\eta\rangle_{|\eta|<0.5}$) at midrapidity for pp and \PbPb collisions at LHC energies are presented in Fig.~\ref{fig:3.11}. Recent measurements of the $\rho^{0}/\pi$ and $\mathrm{K}^{*0}/\mathrm{K}$ ratios show a decreasing trend with increasing multiplicity. These ratios have been investigated using the EPOS4 model, both with and without the UrQMD hadronic afterburner. A significant difference is observed between the predictions of the model with and without UrQMD, with the difference being more pronounced in central \PbPb collisions compared to pp collisions. Furthermore, this difference is more significant for resonances with shorter lifetimes compared to those with longer lifetimes. The EPOS4 model, with the UrQMD hadronic afterburner enabled, successfully captures the features and qualitatively describes the behavior observed in the data.
The \kstar and \phim resonances are ideal candidates for these ratios because they have similar masses and contain strange quarks, while differing by lifetimes of an order of magnitude of 10. The ratios are chosen in such a way that the strangeness enhancement on the production yield is canceled out, allowing for a more direct comparison of resonance yields relative to stable hadrons. Suppression in the $\rm{K^{*0}/K}$ ratio occurs because the decay daughters of the resonance undergo re-scattering processes in the hadronic phase, resulting in a modified final yield of the resonance compared to what was originally produced before chemical freeze-out. This suppression becomes more pronounced with increasing system size. The observed suppression in the $\rm{K^{*0}/K}$ ratio, compared to the $\rm{\phi/K}$ ratio in small to large collision systems, supports the idea that the shorter lifetime of \kstar plays a significant role in modifying the resonance yield in the hadronic phase. Similarly, the short-lived $\rm{\rho^{0}/\pi}$ ratio also shows suppression with increasing multiplicity. Observations from both data and model results indicate that re-scattering effects dominate over regeneration in the hadronic phase for short-lived mesonic resonances.

Similarly, Fig.~\ref{fig:3.11} (right) shows the ratios of baryonic resonances to stable hadrons as a function of \dndeta at midrapidity. 
The $\rm{\Xi^{*0}/\Xi^{-}}$  ratio increases with multiplicity, whereas the $\rm{\Lambda^{*}/\Lambda}$ and $\rm{\Sigma^{*\pm}/\Lambda}$ ratios show no significant change with multiplicity in pp collisions. 
In the \PbPb collisions, the $\rm{\Lambda^{*}/\Lambda}$ ratio decreases with increasing multiplicity, while the $\rm{\Xi^{*0}/\Xi^{-}}$ and $\rm{\Sigma^{*\pm}/\Lambda}$ ratios remain largely unaffected, despite the fact that the $\rm{\Sigma^{*\pm}}$ has a shorter lifetime compared to 
the $\rm{\Lambda^{*}}$. 
This behavior may be attributed to the fact that the decay daughters of these resonances undergo regeneration and rescattering through (pseudo)-elastic interactions in the hadronic phase. In the case of 
$\rm{\Sigma^{*\pm}}$, the regeneration and rescattering effects are expected to cancel each other out, resulting in minimal changes in the yield with increasing system size.  In contrast, the decreasing trend of the $\rm{\Lambda^{*}/\Lambda}$ ratio with multiplicity is primarily driven by rescattering effects. However, recent studies reported in Ref.~\cite{Oliinychenko:2021enj} suggest that the mean free path of resonances also plays a significant role in modifying the final yield.

The $\rm{\Xi^{*0}/\Xi^{-}}$ also show a weak dependence with multiplicity due to larger lifetime compared to $\rm{\Lambda^{*}}$. For \delstar resonance, it is observed that \delstar/p shows flat behavior with multiplicity for \PbPb collisions, whereas the same ratios exhibit a decreasing trend for pp collisions.
Based on experimental data and model calculations, baryonic resonances exhibit distinct behaviors in pp and \PbPb collisions. This shows that the modification of the final reconstructed resonance yields in the hadronic phase is not solely determined by the resonance lifetime but also by other factors mentioned already in Section~\ref{sec:3.1}.

\subsection{Lifetime of hadronic phase}
\label{sec:3.5}
The suppression of short-lived resonances is seen in both ALICE data and the EPOS4 model results with UrQMD. The suppression in the resonance to stable hadron yield is caused by the rescattering of decay products of resonances in the hadronic phase. These ratios act as useful tools for estimating the timespan between chemical and kinetic freeze-outs using the exponential decay law, under the following assumptions: \\
i) Negligible regeneration effects. \\
ii) Simultaneous freeze-out for all particle species.
The relation is given by:
\begin{equation}
    [h^{*}/h]_{kinetic} = [h^{*}/h]_{chemical} \, \times \, e^{-\tau/\tau_{h^*}}
    \label{eq:lifetime}
\end{equation}
\begin{figure}[!htbp]
    \centering
        \includegraphics[width=0.48\textwidth]{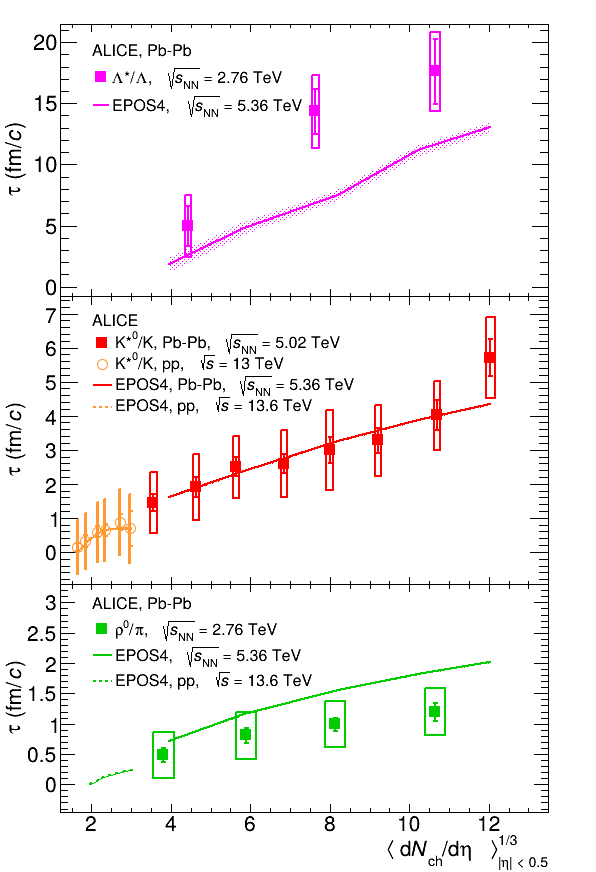}   
    \caption{Lower limit on the lifetime of the hadronic phase between chemical and kinetic freeze-outs in \PbPb collisions across different V0M multiplicity classes, obtained from yield ratios: $\rm{\Lambda^{*}/\Lambda}$ at \snn = 2.76 \TeV (magenta markers)~\cite{ALICE:2018ewo}, $\rm{K}^{*0}/\mathrm{K}$ at \snn = 5.02 \TeV (red markers)~\cite{ALICE:2019xyr}, and $\rho^{0}/\pi$ at \snn = 2.76 \TeV (green markers)~\cite{ALICE:2018qdv}. EPOS4 predictions for \PbPb collisions at \snn = 5.36 \TeV are shown as lines, with shaded areas representing statistical uncertainties. Statistical and systematic uncertainties in the ALICE data are depicted by bars and boxes, respectively. The figure also includes ALICE measurements of $\rm{K}^{*0}/\mathrm{K}$ in pp collisions at $\sqrt{s} = 13$ \TeV (orange markers)~\cite{ALICE:2019etb} and EPOS4 predictions for pp at $\sqrt{s} = 13.6$ \TeV (orange line), highlighting the system size dependence of the hadronic phase lifetime.
}
    \label{fig:3.12} 
\end{figure}
where \([h^{*}/h]_{kinetic}\) denotes the ratio of resonance to stable hadron yield at kinetic freeze-out, \(\tau_{h^*}\) is the lifetime of the resonance particle in its rest frame and \(\tau\) is the time span of the hadronic phase.
Under these assumptions, all resonance particles decaying before kinetic freeze-out are lost due to rescattering of their decay daughters in the hadronic gas, with no regeneration of resonances through elastic scattering. Thus, the estimated time duration is considered as the lower limit for the timespan between chemical and kinetic freeze-outs.

For this calculation, it is further assumed that no hadronic phase forms in pp collisions due to the small system size. Therefore, the yield ratio in minimum bias pp collisions is used as a proxy for  \([h^{*}/h]_{chemical}\). A Lorentz boost factor is also applied in the calculation of $\tau$ because the lifetime of resonance is dilated when transitioning from the rest frame of the resonance to the laboratory reference frame. This factor is approximated as \(\sqrt{1+(\langle \pt \rangle/mc)^2}\), where \(\langle \pt \rangle\) is the average transverse momentum and \(m\) is the rest mass of the resonance particle. The methodology used to estimate $\tau$ is the same as discussed in Ref.~\cite{ALICE:2019xyr}. The results for the estimated duration of the hadronic phase ($\tau$) as a function of ${\langle dN_{ch}/d\eta\rangle}^{1/3}$ from ALICE measurements and EPOS4 with UrQMD  using resonances (\rh, \kstar and \lstar) at LHC energies for pp and \PbPb collisions, are shown in Fig.~\ref{fig:3.12}. 
An increasing trend  in the lifetime of hadronic phase is observed for these resonances in both data and model, corresponding to the decreasing yield ratios (as shown in Fig.~\ref{fig:3.11}) as a function of system size. This trend is expected, as a larger system size at a fixed chemical freeze-out leads to a decrease in the kinetic freeze-out temperature, thereby  a longer timespan exists between chemical and kinetic freeze-outs. This pattern is consistent with the simultaneous blast-wave fits of the identified particle \pt distributions~\cite{ALICE:2019hno}. The timespan of the hadronic phase calculated from the $\rho^0/\pi$, $\mathrm{K}^{*0}/\mathrm{K}$ ratios, are lower than that from the $\rm{\Lambda^{*}/\Lambda}$ ratio. The model predictions for estimation of $\tau$  using the $\rm{\Lambda^{*}/\Lambda}$ ratios underestimate the data, whereas $\rho^{0}/\pi$ overestimate the data. Although the experimental data and model predictions are obtained at slightly different energies, the comparison reveals only minor differences. Since particle production is primarily driven by charged-particle multiplicity, the dependence on collision energy is expected to be weak or negligible. However, the model predictions for $\mathrm{K}^{*0}/\mathrm{K}$ ratio are in good agreement with the data. Contrary to the expectation of a common hadronic phase duration for all resonances, different values are found for the $\rho^{0}$, \kstar, and \lstar yields, with longer-lived resonances showing longer timescales. The observed different values in the time duration of the hadronic phase across different species may be attributed to additional factors beyond rescattering, which also contribute to the final modification of resonance yields. These factors are not accounted for in the simple exponential decay model, as described in Eq.~\ref{eq:lifetime}.
If regeneration is significant, the estimated duration from the yield ratios represents the timescale between delayed resonance production (caused by regeneration) and kinetic freeze-out. This duration reflects a lower limit for the actual hadronic-phase duration. To find the exact duration, the delay from regeneration needs to be added to this estimate.  
Recent calculations using a hybrid approach, where hydrodynamic evolution is followed by the SMASH hadronic transport model, reported that the mean free path of resonances also plays an important role in their final yields~\cite{Oliinychenko:2021enj}. Similar studies in particularly at high multiplicity pp collisions also play huge interest for better understanding of the microscopic origin of resonance suppression. The EPOS4 model results for \kstar with UrQMD in high-multiplicity pp collisions also show a nonzero hadronic phase duration, approximately matching the value in peripheral \PbPb collisions, thus highlighting the system size dependence of the hadronic phase lifetime, as shown in Fig.~\ref{fig:3.12}.

\subsection{Strangeness enhancement}
\label{sec:3.6}
\begin{figure*}[!htbp]
\centering
 \includegraphics[width=1.0\textwidth]{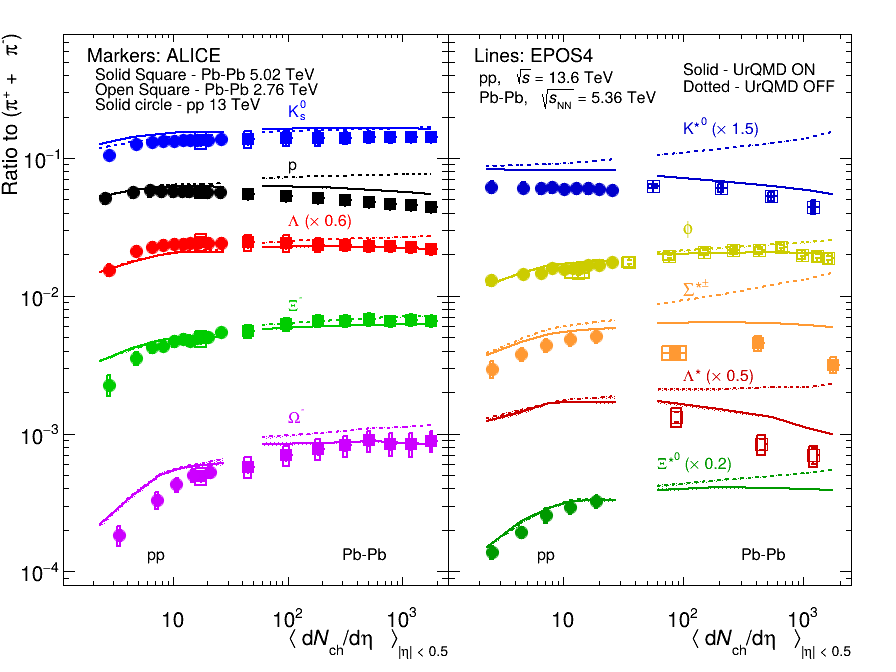}
\caption{Left panel shows the ratios of proton, strange and multi-strange hadrons while the right panel shows the ratios of hadronic resonances, normalized to pions yield. Different markers represent ALICE results in pp collisions at \snn = 13 \TeV (solid circles) \cite{ALICE:2019avo, ALICE:2020nkc, ALICE:2020jsh, ALICE:2019etb, ALICE:2023egx} and \PbPb collisions at \snn = 2.76 \TeV (open squares) \cite{ALICE:2014jbq, ALICE:2018ewo} and \snn = 5.02 \TeV (solid squares) \cite{ALICE:2022zuc, ALICE:2019hno}. The statistical and systematic uncertainties in the data are represented by bars and boxes, respectively. The lines represent predictions from EPOS4 in pp collisions at \snn = 13.6 \TeV and in \PbPb collisions at \snn = 5.36 \TeV with UrQMD (solid) and without UrQMD (dotted). The shaded area indicates the statistical uncertainty in the model calculations.}
\label{fig:3.111}       
\end{figure*}

The relative yield ratios of various hadrons to pions serve as a key observable to understand the origin of strangeness production and the hadronization mechanism. Fig.~\ref{fig:3.111} shows the yield ratios of protons, \kzero and multi-strange baryons (left), along with hadronic resonances (right), normalized to pions as a function of charged-particle multiplicity for pp to \PbPb collisions at LHC energies. The results of EPOS4 with and without UrQMD are represented by solid and dotted lines, respectively, and are compared with the available measurements from the ALICE Collaboration~\cite{ALICE:2019avo, ALICE:2020nkc, ALICE:2020jsh, ALICE:2019etb, ALICE:2023egx, ALICE:2014jbq, ALICE:2018ewo, ALICE:2022zuc, ALICE:2019hno}. The yield ratios exhibit a smooth evolution with multiplicity from pp to \PbPb collisions, whereas the \meanpt shows a non-monotonic pattern with multiplicity, as seen in Fig.~\ref{fig:3.8} and also explained in Section~\ref{sec:3.3}.

\begin{figure}[!htbp]
\centering
\includegraphics[width=0.48\textwidth]{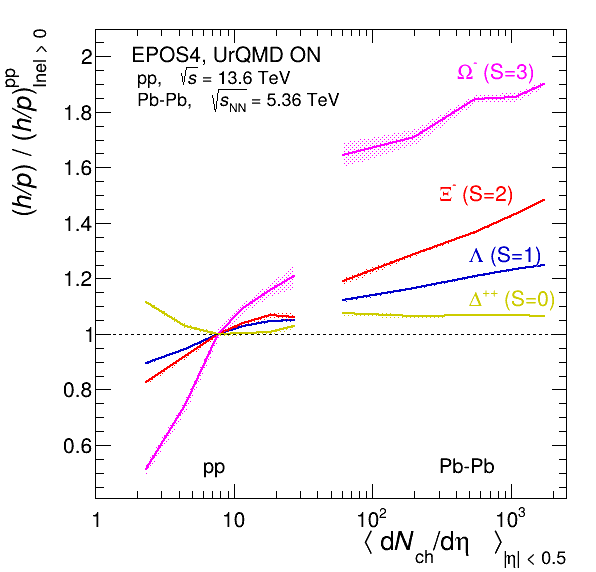}
\caption{Particle yield ratios to proton, normalized to the values measured inelastic pp collisions, as predicted by the EPOS4 model with UrQMD. The results are shown for \delstar and multi-strange baryons in \PbPb collisions at \snn = 5.36 \TeV and in  pp collisions at \s =13.6 \TeV. The bands in the measurements represent statistical uncertainty.}
\label{fig:3.112}       
\end{figure}
 The observed enhancement in the production of strange hadrons relative to pions is primarily driven by their strange-quark content rather than their mass. This effect becomes more pronounced for hadrons with higher strangeness. The change in the slope of the ratios with multiplicity in  the pp collisions shows a steeper rise compared to the \PbPb collisions. The same ratios exhibit a saturation trend for hadrons that contain strange quarks at higher multiplicities for \PbPb collisions. This behavior is understood to be the case that their production reaches a chemical equilibrium within the thermalized environment, where the yields of strange and multi-strange hadrons become independent of further enhancement with increases in multiplicity at LHC energies.
 The EPOS4 model with MCE framework, which reproduces the observed behavior seen in the experimental measurements. However, when the size of the  system gets very small, one gets suppression in the production of heavy particles due to the microcanonical treatment of strange particle production, which imposes constraints on energy and flavor conservation. This effect becomes more pronounced with increasing particle mass and is more significant for $\Omega$ and $\Xi$ baryons in pp collisions. In \PbPb collisions, the model results with and without UrQMD show differences for protons and multi-strange hadrons, highlighting the role of hadronic phase effects. In contrast, no significant difference is observed for pp collisions. Moreover, the proton-to-pion yield ratios from EPOS4 simulations with UrQMD show a decreasing trend with multiplicity in \PbPb collisions, in qualitative agreement with experimental observations, but the model slightly overestimates the measured values.
One of the possible explanations for this suppression is baryon-antibaryon annihilation that occurs in the hadronic phase within the high-multiplicity, high-density environment of \PbPb collisions, which reduces the overall proton yield relative to pions. However, the same ratios for $\rm{\Lambda}$ show only a weak decrease, while for heavier strange baryons such as $\rm{\Xi^{-}}$ and $\rm{\Omega^{-}}$ there is no decreasing trend with multiplicity.

To further understand the behavior, the double yield ratios of $\delstar$, $\rm{\Lambda}$, $\rm{\Xi^{-}}$, and $\rm{\Omega^{-}}$ to protons have been calculated. These ratios are normalized to their respective values measured in inelastic pp collisions. Calculations were performed as a function of charged particle multiplicity using EPOS4 with the UrQMD hadronic afterburner. The results, shown in Fig.~\ref{fig:3.112}, demonstrate that the multiplicity-dependent enhancement relative to inelastic pp collisions follows a hierarchy determined by  strangeness content of hadrons. To follow up on the earlier discussion about the decreasing trend in proton-to-pion ratios, another non-strange baryon, $\delstar$, is chosen. Since $\delstar$ contains no strange quarks and shows no hadronic phase effects (as discussed in Fig.~\ref{fig:3.11}), it serves as a useful candidate. The double yield ratios for $\delstar$ exhibit a flat trend with multiplicity in \PbPb collisions, reflecting that baryon–antibaryon annihilation affects $\delstar$ in a similar manner as  protons in the hadronic phase. In contrast, the $\rm{\Lambda}$-to-pion ratio shows only a weak dependence on multiplicity, while $\rm{\Xi^{-}}$ and $\rm{\Omega^{-}}$ exhibit no decreasing trend, even in the most central \PbPb collisions. According to the strangeness enhancement picture, the yield ratios reach a saturation trend at higher multiplicities. The slight decrease in the $\rm{\Lambda}$-to-pion ratio at the most central collisions could be attributed to additional contributions from baryon–antibaryon annihilation during the hadronic phase. This modifies strange-baryon productions and acts opposite to the  strangeness enhancement scenario. Therefore, the net effect on baryon-to-pion ratios depends on the interplay between strangeness enhancement and baryon–antibaryon annihilation. Strangeness enhancement is found to be the dominant mechanism for strange baryon production, whereas the contribution 
from baryon–antibaryon annihilation is  negligible, particularly for heavier strange baryons, due to their lower abundance resulting from larger mass and reduced production probability. Moreover, the heavier multi-strange baryons are expected to freeze out earlier, reducing their interactions in the hadronic phase and lowering the likelihood of annihilation compared to protons~\cite{PhysRevC.97.024913}. Future experimental precision measurements of these ratios at the LHC will provide deeper insight, helping to distinguish between the effects of strangeness enhancement and baryon-antibaryon annihilation.

Furthermore, the ratios of short-lived hadronic resonances ($\rho$, $\kstar$, and $\lstar$) to pions show a decreasing trend with multiplicity in Pb–Pb collisions. A similar trend is observed in the experimental measurements, and the trend is also reproduced by EPOS4 with UrQMD. This behavior is understood as the rescattering effect dominating over strangeness enhancement for short-lived resonances. Similarly, for the long-lived $\phi$-meson, a weak enhancement trend is observed, except in the most central Pb–Pb collisions, where measurements reveal a decreasing trend, which is reproduced by UrQMD. Future measurements will provide further constraints to better understand this behavior.
The relative yield ratios of hadrons to pions are determined by the combined contributions of various processes, including strangeness production, baryon–antibaryon annihilation, and hadronic phase effects, with the significance of each process varying based on the specific hadron species.

\section{Summary}
\label{sec:4}
In this work, we present the study of various hadronic resonances to explore production dynamics and their interactions, along with other stable strange and non-strange hadrons at midrapidity in pp and \PbPb collisions at \s = 13.6 TeV and \snn = 5.36 TeV, using the EPOS4 simulation. The results are new findings with the latest EPOS4 model, extending such studies for the first time to high-multiplicity pp collisions and the highest energies for \PbPb collisions at the LHC. 
The EPOS4, with UrQMD as the hadronic afterburner, is a well-suited model to describe experimental measurements of resonance production in high-energy collisions. It reproduces the qualitative trend of resonance-to-ground-state particle yield ratios as a function of system size, from pp to \PbPb\ collisions. The ratios of shorter-lived resonances (\rh, \kstar, \lstar) are suppressed, while \xistar\ exhibits a weak suppression, and \sigmstar\ and \delstar\ remain unaffected. This suppression is not observed in the \phim  meson due to its longer lifetime compared to other resonances.  
Precision studies of heavier baryonic resonances (\xistar, \sigmstar) and predictions for \delstar await confirmation through future measurements in heavy-ion collisions at the LHC. Similar ratios generate strong interest in small systems, since model calculations of the $\rho/\pi$ and $K^{*0}/K$ ratios show a decreasing trend, with stronger suppression at high multiplicity. This trend is consistent with the measured $K^{*0}/K$ ratio in pp collisions by the ALICE Collaboration. The lower limit of the time duration ($\tau$) of the hadronic phase can be estimated using the yield ratios of $\rho/\pi$, $K^{*0}/K$, and $\lstar/\Lambda$, and it increases with the system size. A non-zero time duration ($\sim$0.5--1 fm/$c$) is found in high-multiplicity pp collisions. The suppression in these ratios arises from modifications of resonance yields in the hadronic phase, where rescattering processes dominate over regeneration. This effect is more pronounced in the low-\pt region and is particularly significant for short-lived resonances.

Enabling UrQMD improves the description of the \meanpt values for various hadronic resonances and protons. In central \PbPb collisions, the mass ordering of \meanpt for particles with similar masses (e.g., p and \phim) aligns with hydrodynamic predictions, but this ordering breaks down in peripheral \PbPb and pp collisions. The \meanpt as a function of reduced mass (scaled by the number of valence quarks) shows a linearly increasing trend. The baryons and mesons form distinct groups that are parallel to each other. For short-lived resonances such as \rh and \kstar, deviations from the linear trend are observed in high-multiplicity pp collisions and  more significant for the central \PbPb collisions. This observation is a consequence of rescattering effects in the hadronic phase, which leads to an increase in the $\langle p_{\mathrm{T}} \rangle$ values of short-lived resonances. Moreover, for stable hadrons like proton and  \lmb, the \meanpt values differ significantly with and without UrQMD, particularly in central \PbPb collisions, their spectral shapes and \meanpt values are influenced by the decays of higher-mass resonances.

The baryon-to-meson ratios in central and peripheral \PbPb collisions at \snn = 5.36 \TeV have been measured. This confirms that the radial flow plays a dominant role in enhancing the $\rm{\Lambda^{*}/\kzero}$, and $\rm{\Sigma^{*\pm}/\kzero}$ ratios in the intermediate \pt region. Additionally, the $\rm{p/\phi}$ and $\rm{p/K^{*0}}$ ratios remain nearly flat in the intermediate \pt. At high \pt, all the ratios exhibit the similar decreasing trend that indicates the common fragmentation process influencing their production. To understand the microscopic origin of the strangeness production, the strange and multi-strange hadrons along with resonances are investigated. The relative yield ratios of hadrons to pions smoothly increase with multiplicity, with the rate of increase depending on the strange quark content of the hadrons. 
However, the ratios reach the saturation trend at higher multiplicities 
($\langle\mathrm{d}N_\mathrm{ch}/\mathrm{d}\eta\rangle_{|\eta|<0.5} >  100$), consistent with expectations that strangeness production approaches the grand-canonical limit at large system size and high density, where canonical effects become negligible.
The p/$\pi$ ratios exhibit the decreasing trend with increasing multiplicity, the behavior that becomes more pronounced at higher multiplicities for \PbPb collisions. This trend is observed in both experimental data and model predictions. The suppression at higher densities is primarily driven by baryon-antibaryon annihilation for non-strange hadrons ($p$) whereas this effect is less pronounced for heavier multi-strange baryons (\lmb, $\rm{\Omega}$, and $\rm{\Xi}$). The \pt spectra and yields of various hadrons, including resonances, have been studied to understand the effects of the late-stage hadronic phase by EPOS4 with UrQMD. 

From these studies, it is concluded that hadronic phase effects, such as rescattering and regeneration, play an important role in the modification of resonance yields relative to stable hadrons. For mesonic resonances ($\rho$, $\kstar$, $\phi$), the observed suppression pattern follows their vacuum lifetimes, consistent with expectations from rescattering in the hadronic medium. In contrast, baryonic resonances exhibit a more complex behavior, while the short-lived $\Sigma^{*}$ shows no significant suppression, the $\Lambda^{*}$ displays a stronger suppression than the $\kstar$ meson. This could be because regeneration contributions depend on the decay products and their interaction cross sections in the hadronic phase. The final modification of resonance yields cannot be attributed to a single factor, but rather results from the combined effects of decay product cross sections, lifetime of resonance, the timespan of system created, and assumption of single  freeze-out temperature for all hadrons.

Further, differential studies, such as resonance flow and correlations with jets, will provide insights into the hadronic phase created in high-energy collisions at LHC energies. A comprehensive understanding of the hadronic phase, even in small collision systems, is therefore essential for accurately interpreting observables that reflect the underlying production dynamics. Future measurements of resonances such as the $\Delta^{*}$, $f_{0}(980)$, $\Lambda(1405)$, $N(1535)$, and $K_{1}(1270)$ will provide additional input on the late-stage interactions in high-energy collisions.

\section{Acknowledgement}
We sincerely thank K. Werner for his invaluable contributions through insightful discussions and for providing critical technical details related to the EPOS4 model. The authors at University of Jammu gratefully acknowledge the Department of Science and Technology (DST), Government of India, for their financial support in doing this work. We also extend our gratitude to the CERN Grid computing centre for providing the computational resources required for this work.
\dots.
\bibliography{mybib1}  

@PREAMBLE{
 "\providecommand{\noopsort}[1]{}" 
 # "\providecommand{\singleletter}[1]{#1}%" 
}

@article{Harris:1996zx,
    author = "Harris, John W. and Muller, Berndt",
    title = "{The Search for the quark - gluon plasma}",
    eprint = "hep-ph/9602235",
    archivePrefix = "arXiv",
    reportNumber = "DUKE-TH-96-105",
    doi = "10.1146/annurev.nucl.46.1.71",
    journal = "Ann. Rev. Nucl. Part. Sci.",
    volume = "46",
    pages = "71--107",
    year = "1996"
}

@article{ALICE:2019xyr,
    author = "Acharya, Shreyasi and others",
    collaboration = "ALICE",
    title = "{Evidence of rescattering effect in Pb-Pb collisions at the LHC through production of $\rm{K}^{*}(892)^{0}$ and $\phi(1020)$ mesons}",
    eprint = "1910.14419",
    archivePrefix = "arXiv",
    primaryClass = "nucl-ex",
    reportNumber = "CERN-EP-2019-249",
    doi = "10.1016/j.physletb.2020.135225",
    journal = "Phys. Lett. B",
    volume = "802",
    pages = "135225",
    year = "2020"
}

@article{Knospe:2015nva,
    author = "Knospe, A. G. and Markert, C. and Werner, K. and Steinheimer, J. and Bleicher, M.",
    title = "{Hadronic resonance production and interaction in partonic and hadronic matter in the EPOS3 model with and without the hadronic afterburner UrQMD}",
    eprint = "1509.07895",
    archivePrefix = "arXiv",
    primaryClass = "nucl-th",
    doi = "10.1103/PhysRevC.93.014911",
    journal = "Phys. Rev. C",
    volume = "93",
    number = "1",
    pages = "014911",
    year = "2016"
}

@article{ALICE:2018qdv,
    author = "Acharya, Shreyasi and others",
    collaboration = "ALICE",
    title = "{Production of the $\rho$(770)${^{0}}$ meson in pp and Pb-Pb collisions at $\sqrt{s_{\rm NN}}$ = 2.76 TeV}",
    eprint = "1805.04365",
    archivePrefix = "arXiv",
    primaryClass = "nucl-ex",
    reportNumber = "CERN-EP-2018-106",
    doi = "10.1103/PhysRevC.99.064901",
    journal = "Phys. Rev. C",
    volume = "99",
    number = "6",
    pages = "064901",
    year = "2019"
}

@article{ALICE:2021ptz,
    author = "Acharya, Shreyasi and others",
    collaboration = "ALICE",
    title = "{Production of K$^{*}(892)^{0}$ and $\phi(1020)$ in pp and Pb-Pb collisions at $\sqrt{s_{\rm NN}} = 5.02$ TeV}",
    eprint = "2106.13113",
    archivePrefix = "arXiv",
    primaryClass = "nucl-ex",
    reportNumber = "CERN-EP-2021-101",
    doi = "10.1103/PhysRevC.106.034907",
    journal = "Phys. Rev. C",
    volume = "106",
    number = "3",
    pages = "034907",
    year = "2022"
}

@article{ALICE:2022zuc,
    author = "Acharya, Shreyasi and others",
    collaboration = "ALICE",
    title = "{$\Sigma (1385)^{\pm }$ resonance production in Pb\textendash{}Pb collisions at $\sqrt{s_{\textrm{NN}}}~=~5.02$~TeV}",
    eprint = "2205.13998",
    archivePrefix = "arXiv",
    primaryClass = "nucl-ex",
    reportNumber = "CERN-EP-2022-104",
    doi = "10.1140/epjc/s10052-023-11475-1",
    journal = "Eur. Phys. J. C",
    volume = "83",
    number = "5",
    pages = "351",
    year = "2023"
}

@article{Gaudichet:2003jr,
    author = "Gaudichet, Ludovic",
    editor = "Bass, S. A. and Muller, Berndt and Stephans, G. S. F. and Ullrich, T.",
    collaboration = "STAR",
    title = "{Lambda(1520) and Sigma(1385) resonance production in Au+Au and p+p collisions at RHIC (S(NN)**1/2 = 200-GeV)}",
    eprint = "nucl-ex/0307013",
    archivePrefix = "arXiv",
    doi = "10.1088/0954-3899/30/1/067",
    journal = "J. Phys. G",
    volume = "30",
    pages = "S549--S555",
    year = "2004"
}

@article{ALICE:2018ewo,
    author = "Acharya, Shreyasi and others",
    collaboration = "ALICE",
    title = "{Suppression of $\Lambda(1520)$ resonance production in central Pb-Pb collisions at $\sqrt{s_{\rm NN}}$ = 2.76 TeV}",
    eprint = "1805.04361",
    archivePrefix = "arXiv",
    primaryClass = "nucl-ex",
    reportNumber = "CERN-EP-2018-116",
    doi = "10.1103/PhysRevC.99.024905",
    journal = "Phys. Rev. C",
    volume = "99",
    pages = "024905",
    year = "2019"
}

@article{ALICE:2019smg,
    author = "Acharya, S. and others",
    collaboration = "ALICE",
    title = "{Measurement of $\Lambda$(1520) production in pp collisions at $\sqrt{s}$ = 7 TeV and p-Pb collisions at $\sqrt{s_{\rm{NN}}}$ = 5.02 TeV}",
    eprint = "1909.00486",
    archivePrefix = "arXiv",
    primaryClass = "nucl-ex",
    reportNumber = "CERN-EP-2019-178",
    doi = "10.1140/epjc/s10052-020-7687-2",
    journal = "Eur. Phys. J. C",
    volume = "80",
    number = "2",
    pages = "160",
    year = "2020"
}

@article{ALICE:2021rpa,
    author = "Acharya, Shreyasi and others",
    collaboration = "ALICE",
    title = "{K*(892)0 and \ensuremath{\phi}(1020) production in p-Pb collisions at sNN=8.16~TeV}",
    eprint = "2110.10042",
    archivePrefix = "arXiv",
    primaryClass = "nucl-ex",
    reportNumber = "CERN-EP-2021-200",
    doi = "10.1103/PhysRevC.107.055201",
    journal = "Phys. Rev. C",
    volume = "107",
    number = "5",
    pages = "055201",
    year = "2023"
}

@article{Werner:2013tya,
    author = "Werner, K. and Guiot, B. and Karpenko, Iu. and Pierog, T.",
    title = "{Analysing radial flow features in p-Pb and p-p collisions at several TeV by studying identified particle production in EPOS3}",
    eprint = "1312.1233",
    archivePrefix = "arXiv",
    primaryClass = "nucl-th",
    doi = "10.1103/PhysRevC.89.064903",
    journal = "Phys. Rev. C",
    volume = "89",
    number = "6",
    pages = "064903",
    year = "2014"
}

@article{ALICE:2019avo,
    author = "Acharya, Shreyasi and others",
    collaboration = "ALICE",
    title = "{Multiplicity dependence of (multi-)strange hadron production in proton-proton collisions at $\sqrt{s}$ = 13 TeV}",
    eprint = "1908.01861",
    archivePrefix = "arXiv",
    primaryClass = "nucl-ex",
    reportNumber = "CERN-EP-2019-168",
    doi = "10.1140/epjc/s10052-020-7673-8",
    journal = "Eur. Phys. J. C",
    volume = "80",
    number = "2",
    pages = "167",
    year = "2020"
}

@article{ALICE:2019hno,
    author = "Acharya, Shreyasi and others",
    collaboration = "ALICE",
    title = "{Production of charged pions, kaons, and (anti-)protons in Pb-Pb and inelastic $pp$ collisions at $\sqrt {s_{NN}}$ = 5.02 TeV}",
    eprint = "1910.07678",
    archivePrefix = "arXiv",
    primaryClass = "nucl-ex",
    reportNumber = "CERN-EP-2019-208",
    doi = "10.1103/PhysRevC.101.044907",
    journal = "Phys. Rev. C",
    volume = "101",
    number = "4",
    pages = "044907",
    year = "2020"
}

@article{ALICE:2014sbx,
    author = "Abelev, Betty Bezverkhny and others",
    collaboration = "ALICE",
    title = "{Performance of the ALICE Experiment at the CERN LHC}",
    eprint = "1402.4476",
    archivePrefix = "arXiv",
    primaryClass = "nucl-ex",
    reportNumber = "CERN-PH-EP-2014-031",
    doi = "10.1142/S0217751X14300440",
    journal = "Int. J. Mod. Phys. A",
    volume = "29",
    pages = "1430044",
    year = "2014"
}

@article{Werner:2023jps,
    author = "Werner, K.",
    title = "{Core-corona procedure and microcanonical hadronization to understand strangeness enhancement in proton-proton and heavy ion collisions in the EPOS4 framework}",
    eprint = "2306.10277",
    archivePrefix = "arXiv",
    primaryClass = "hep-ph",
    doi = "10.1103/PhysRevC.109.014910",
    journal = "Phys. Rev. C",
    volume = "109",
    number = "1",
    pages = "014910",
    year = "2024"
}

@article{Pierog:2013ria,
    author = "Pierog, T. and Karpenko, Iu. and Katzy, J. M. and Yatsenko, E. and Werner, K.",
    title = "{EPOS LHC: Test of collective hadronization with data measured at the CERN Large Hadron Collider}",
    eprint = "1306.0121",
    archivePrefix = "arXiv",
    primaryClass = "hep-ph",
    reportNumber = "DESY-13-125",
    doi = "10.1103/PhysRevC.92.034906",
    journal = "Phys. Rev. C",
    volume = "92",
    number = "3",
    pages = "034906",
    year = "2015"
}

@article{Gribov:1967vfb,
    author = "Gribov, V. N.",
    title = "{A REGGEON DIAGRAM TECHNIQUE}",
    journal = "Zh. Eksp. Teor. Fiz.",
    volume = "53",
    pages = "654--672",
    year = "1967"
}

@article{Gribov:1968jf,
    author = "Gribov, V. N.",
    title = "{Glauber corrections and the interaction between high-energy hadrons and nuclei}",
    journal = "Sov. Phys. JETP",
    volume = "29",
    pages = "483--487",
    year = "1969"
}

@article{Gribov:1972ri,
    author = "Gribov, V. N. and Lipatov, L. N.",
    title = "{Deep inelastic e p scattering in perturbation theory}",
    reportNumber = "IPTI-381-71",
    journal = "Sov. J. Nucl. Phys.",
    volume = "15",
    pages = "438--450",
    year = "1972"
}

@article{Altarelli:1977zs,
    author = "Altarelli, Guido and Parisi, G.",
    title = "{Asymptotic Freedom in Parton Language}",
    reportNumber = "LPTENS-77-6",
    doi = "10.1016/0550-3213(77)90384-4",
    journal = "Nucl. Phys. B",
    volume = "126",
    pages = "298--318",
    year = "1977"
}

@article{Dokshitzer:1977sg,
    author = "Dokshitzer, Yuri L.",
    title = "{Calculation of the Structure Functions for Deep Inelastic Scattering and e+ e- Annihilation by Perturbation Theory in Quantum Chromodynamics.}",
    journal = "Sov. Phys. JETP",
    volume = "46",
    pages = "641--653",
    year = "1977"
}

@article{Abramovsky:1973fm,
    author = "Abramovsky, V. A. and Gribov, V. N. and Kancheli, O. V.",
    title = "{Character of Inclusive Spectra and Fluctuations Produced in Inelastic Processes by Multi - Pomeron Exchange}",
    journal = "Yad. Fiz.",
    volume = "18",
    pages = "595--616",
    year = "1973"
}

@article{Werner:2007bf,
    author = "Werner, Klaus",
    title = "{Core-corona separation in ultra-relativistic heavy ion collisions}",
    eprint = "0704.1270",
    archivePrefix = "arXiv",
    primaryClass = "nucl-th",
    doi = "10.1103/PhysRevLett.98.152301",
    journal = "Phys. Rev. Lett.",
    volume = "98",
    pages = "152301",
    year = "2007"
}

@article{Werner:2010aa,
    author = "Werner, K. and Karpenko, Iu. and Pierog, T. and Bleicher, M. and Mikhailov, K.",
    title = "{Event-by-Event Simulation of the Three-Dimensional Hydrodynamic Evolution from Flux Tube Initial Conditions in Ultrarelativistic Heavy Ion Collisions}",
    eprint = "1004.0805",
    archivePrefix = "arXiv",
    primaryClass = "nucl-th",
    doi = "10.1103/PhysRevC.82.044904",
    journal = "Phys. Rev. C",
    volume = "82",
    pages = "044904",
    year = "2010"
}

@article{Bleicher:1999xi,
    author = "Bleicher, M. and others",
    title = "{Relativistic hadron hadron collisions in the ultrarelativistic quantum molecular dynamics model}",
    eprint = "hep-ph/9909407",
    archivePrefix = "arXiv",
    doi = "10.1088/0954-3899/25/9/308",
    journal = "J. Phys. G",
    volume = "25",
    pages = "1859--1896",
    year = "1999"
}

@article{Gribov:1983ivg,
    author = "Gribov, L. V. and Levin, E. M. and Ryskin, M. G.",
    title = "{Semihard Processes in QCD}",
    doi = "10.1016/0370-1573(83)90022-4",
    journal = "Phys. Rept.",
    volume = "100",
    pages = "1--150",
    year = "1983"
}

@article{McLerran:1993ni,
    author = "McLerran, Larry D. and Venugopalan, Raju",
    title = "{Computing quark and gluon distribution functions for very large nuclei}",
    eprint = "hep-ph/9309289",
    archivePrefix = "arXiv",
    reportNumber = "TPI-MINN-93-44-T, NUC-MINN-93-24-T, HEP-UMN-TH-1220-93",
    doi = "10.1103/PhysRevD.49.2233",
    journal = "Phys. Rev. D",
    volume = "49",
    pages = "2233--2241",
    year = "1994"
}

@article{McLerran:1993ka,
    author = "McLerran, Larry D. and Venugopalan, Raju",
    title = "{Gluon distribution functions for very large nuclei at small transverse momentum}",
    eprint = "hep-ph/9311205",
    archivePrefix = "arXiv",
    reportNumber = "TPI-MINN-93-52-T, NUC-MINN-93-28-T, UMN-TH-1224-93",
    doi = "10.1103/PhysRevD.49.3352",
    journal = "Phys. Rev. D",
    volume = "49",
    pages = "3352--3355",
    year = "1994"
}

@article{Kovner:1995ts,
    author = "Kovner, Alex and McLerran, Larry D. and Weigert, Heribert",
    title = "{Gluon production at high transverse momentum in the McLerran-Venugopalan model of nuclear structure functions}",
    eprint = "hep-ph/9505320",
    archivePrefix = "arXiv",
    reportNumber = "TPI-MINN-95-16-T, NUC-MINN-95-14-T, HEP-MINN-95-1346",
    doi = "10.1103/PhysRevD.52.3809",
    journal = "Phys. Rev. D",
    volume = "52",
    pages = "3809--3814",
    year = "1995"
}

@article{Kovchegov:1996ty,
    author = "Kovchegov, Yuri V.",
    title = "{NonAbelian Weizsacker-Williams field and a two-dimensional effective color charge density for a very large nucleus}",
    eprint = "hep-ph/9605446",
    archivePrefix = "arXiv",
    reportNumber = "CU-TP-753",
    doi = "10.1103/PhysRevD.54.5463",
    journal = "Phys. Rev. D",
    volume = "54",
    pages = "5463--5469",
    year = "1996"
}

@article{ALICE:2019etb,
    author = "Acharya, Shreyasi and others",
    collaboration = "ALICE",
    title = "{Multiplicity dependence of K*(892)0 and \ensuremath{\phi}(1020) production in pp collisions at s=13 TeV}",
    eprint = "1910.14397",
    archivePrefix = "arXiv",
    primaryClass = "nucl-ex",
    reportNumber = "CERN-EP-2019-245",
    doi = "10.1016/j.physletb.2020.135501",
    journal = "Phys. Lett. B",
    volume = "807",
    pages = "135501",
    year = "2020"
}

@article{ALICE:2016sak,
    author = "Adam, Jaroslav and others",
    collaboration = "ALICE",
    title = "{Production of K$^{*}$ (892)$^{0}$ and $\phi $ (1020) in p\textendash{}Pb collisions at $\sqrt{s_{{\text {NN}}}}$ = 5.02 TeV}",
    eprint = "1601.07868",
    archivePrefix = "arXiv",
    primaryClass = "nucl-ex",
    reportNumber = "CERN-PH-EP-2015-326",
    doi = "10.1140/epjc/s10052-016-4088-7",
    journal = "Eur. Phys. J. C",
    volume = "76",
    number = "5",
    pages = "245",
    year = "2016"
}

@article{STAR:2004bgh,
    author = "Adams, J. and others",
    collaboration = "STAR",
    title = "{K(892)* resonance production in Au+Au and p+p collisions at s(NN)**(1/2) = 200-GeV at STAR}",
    eprint = "nucl-ex/0412019",
    archivePrefix = "arXiv",
    doi = "10.1103/PhysRevC.71.064902",
    journal = "Phys. Rev. C",
    volume = "71",
    pages = "064902",
    year = "2005"
}

@article{ALICE:2023egx,
    author = "Acharya, Shreyasi and others",
    collaboration = "ALICE",
    title = "{Multiplicity-dependent production of \ensuremath{\Sigma}(1385)$^{±}$ and \ensuremath{\Xi}(1530)$^{0}$ in pp collisions at $ \sqrt{s} $ = 13 TeV}",
    eprint = "2308.16116",
    archivePrefix = "arXiv",
    primaryClass = "nucl-ex",
    reportNumber = "CERN-EP-2023-172",
    doi = "10.1007/JHEP05(2024)317",
    journal = "JHEP",
    volume = "05",
    pages = "317",
    year = "2024"
}

@article{ALICE:2020nkc,
    author = "Acharya, Shreyasi and others",
    collaboration = "ALICE",
    title = "{Multiplicity dependence of $\pi $, K, and p production in pp collisions at $\sqrt{s} = 13$ TeV}",
    eprint = "2003.02394",
    archivePrefix = "arXiv",
    primaryClass = "nucl-ex",
    reportNumber = "CERN-EP-2020-024",
    doi = "10.1140/epjc/s10052-020-8125-1",
    journal = "Eur. Phys. J. C",
    volume = "80",
    number = "8",
    pages = "693",
    year = "2020"
}

@article{ALICE:2020jsh,
    author = "Acharya, Shreyasi and others",
    collaboration = "ALICE",
    title = "{Production of light-flavor hadrons in pp collisions at $\sqrt{s}~=~7\text { and }\sqrt{s} = 13 \, \text { TeV} $}",
    eprint = "2005.11120",
    archivePrefix = "arXiv",
    primaryClass = "nucl-ex",
    reportNumber = "CERN-EP-2020-059",
    doi = "10.1140/epjc/s10052-020-08690-5",
    journal = "Eur. Phys. J. C",
    volume = "81",
    number = "3",
    pages = "256",
    year = "2021"
}

@article{ALICE:2014jbq,
    author = "Abelev, Betty Bezverkhny and others",
    collaboration = "ALICE",
    title = "{$K^*(892)^0$ and $ϕ(1020)$ production in Pb-Pb collisions at $\sqrt{s{NN}}$ = 2.76 TeV}",
    eprint = "1404.0495",
    archivePrefix = "arXiv",
    primaryClass = "nucl-ex",
    reportNumber = "CERN-PH-EP-2014-060",
    doi = "10.1103/PhysRevC.91.024609",
    journal = "Phys. Rev. C",
    volume = "91",
    pages = "024609",
    year = "2015"
}

@article{Oliinychenko:2021enj,
    author        = "Oliinychenko, Dmytro and Shen, Chun",
    title         = "{Resonance production in Pb-Pb collisions at 5.02 TeV via hydrodynamics and hadronic afterburner}",
    eprint        = "2105.07539",
    archivePrefix = "arXiv",
    primaryClass  = "hep-ph",
    journal       = "arXiv preprint",
    month         = may,
    year          = "2021",
    note          = "[hep-ph]"
}

@article{STAR:2006vhb,
    author = "Abelev, B. I. and others",
    collaboration = "STAR",
    title = "{Strange baryon resonance production in s(NN)**(1/2) = 200-GeV p+p and Au+Au collisions}",
    eprint = "nucl-ex/0604019",
    archivePrefix = "arXiv",
    doi = "10.1103/PhysRevLett.97.132301",
    journal = "Phys. Rev. Lett.",
    volume = "97",
    pages = "132301",
    year = "2006"
}

@article{Lin:2004en,
    author = "Lin, Zi-Wei and Ko, Che Ming and Li, Bao-An and Zhang, Bin and Pal, Subrata",
    title = "{A Multi-phase transport model for relativistic heavy ion collisions}",
    eprint = "nucl-th/0411110",
    archivePrefix = "arXiv",
    doi = "10.1103/PhysRevC.72.064901",
    journal = "Phys. Rev. C",
    volume = "72",
    pages = "064901",
    year = "2005"
}

@article{Motornenko:2019jha,
    author = "Motornenko, Anton and Vovchenko, Volodymyr and Greiner, Carsten and Stoecker, Horst",
    title = "{Kinetic freeze-out temperature from yields of short-lived resonances}",
    eprint = "1908.11730",
    archivePrefix = "arXiv",
    primaryClass = "hep-ph",
    doi = "10.1103/PhysRevC.102.024909",
    journal = "Phys. Rev. C",
    volume = "102",
    number = "2",
    pages = "024909",
    year = "2020"
}

@article{Knospe:2021jgt,
    author = "Knospe, A. G. and Markert, C. and Werner, K. and Steinheimer, J. and Bleicher, M.",
    title = "{Hadronic resonance production and interaction in p-Pb collisions at LHC energies in EPOS3}",
    eprint = "2102.06797",
    archivePrefix = "arXiv",
    primaryClass = "nucl-th",
    doi = "10.1103/PhysRevC.104.054907",
    journal = "Phys. Rev. C",
    volume = "104",
    number = "5",
    pages = "054907",
    year = "2021"
}

@article{Fries:2003vb,
    author = "Fries, R. J. and Muller, Berndt and Nonaka, C. and Bass, S. A.",
    title = "{Hadronization in heavy ion collisions: Recombination and fragmentation of partons}",
    eprint = "nucl-th/0301087",
    archivePrefix = "arXiv",
    reportNumber = "DUKE-TH-03-233",
    doi = "10.1103/PhysRevLett.90.202303",
    journal = "Phys. Rev. Lett.",
    volume = "90",
    pages = "202303",
    year = "2003"
}

@article{ALICE:2022wpn,
    author = "Acharya, Shreyasi and others",
    collaboration = "ALICE",
    title = "{The ALICE experiment: a journey through QCD}",
    eprint = "2211.04384",
    archivePrefix = "arXiv",
    primaryClass = "nucl-ex",
    reportNumber = "CERN-EP-2022-227",
    doi = "10.1140/epjc/s10052-024-12935-y",
    journal = "Eur. Phys. J. C",
    volume = "84",
    number = "8",
    pages = "813",
    year = "2024"
}

@article{PhysRevC.97.024913,
  title = {Baryon-antibaryon annihilation and reproduction in relativistic heavy-ion collisions},
  author = {Seifert, E. and Cassing, W.},
  journal = {Phys. Rev. C},
  volume = {97},
  issue = {2},
  pages = {024913},
  numpages = {17},
  year = {2018},
  month = {Feb},
  publisher = {American Physical Society},
  doi = {10.1103/PhysRevC.97.024913},
  url = {https://link.aps.org/doi/10.1103/PhysRevC.97.024913}
}

\end{document}